\documentclass[3p,preprint,12pt]{elsarticle}
\makeatletter\if@twocolumn\PassOptionsToPackage{switch}{lineno}\else\fi\makeatother

\usepackage{tabulary,xcolor}
\usepackage{amsfonts,amsmath,amssymb}
\usepackage{geometry}
\usepackage{booktabs}
\usepackage[T1]{fontenc}
\makeatletter
\let\save@ps@pprintTitle\ps@pprintTitle
\def\ps@pprintTitle{\save@ps@pprintTitle\gdef\@oddfoot{\footnotesize\itshape \null\hfill\today}}
\def\hlinewd#1{%
  \noalign{\ifnum0=`}\fi\hrule \@height #1%
  \futurelet\reserved@a\@xhline}

\AtBeginDocument{\ifNAT@numbers \biboptions{sort&compress}\fi}

\makeatother

  \renewenvironment{abstract}{\global\setbox\absbox=\vbox\bgroup
    \hsize=\textwidth%
  \noindent\unskip\textbf{}
   \par\medskip\noindent\unskip\ignorespaces}
   {\egroup}

\usepackage{ifluatex}
\ifluatex
\usepackage{fontspec}
\defaultfontfeatures{Ligatures=TeX}
\usepackage[]{unicode-math}
\unimathsetup{math-style=TeX}
\else 
\usepackage[utf8]{inputenc}
\fi 
\ifluatex\else\usepackage{stmaryrd}\fi
\usepackage{graphicx}   % 图形包，插入图片
\numberwithin{figure}{section}  % 让图号变为 1.1, 1.2 ...
\usepackage{caption}

\usepackage{url,multirow,morefloats,floatflt,cancel,tfrupee}
\makeatletter

\AtBeginDocument{\@ifpackageloaded{textcomp}{}{\usepackage{textcomp}}}
\makeatother
\usepackage{colortbl}
\usepackage{xcolor}
\usepackage{pifont}
\usepackage[nointegrals]{wasysym}
\makeatletter

\def\mcWidth#1{\csname TY@F#1\endcsname+\tabcolsep}

\def\cAlignHack{\rightskip\@flushglue\leftskip\@flushglue\parindent\z@\parfillskip\z@skip}
\def\rAlignHack{\rightskip\z@skip\leftskip\@flushglue \parindent\z@\parfillskip\z@skip}

\@ifundefined{etal}{}{}

\usepackage{ifxetex}
\ifxetex\else\if@twocolumn\@ifpackageloaded{stfloats}{}{\usepackage{dblfloatfix}}\fi\fi

\AtBeginDocument{
\expandafter\ifx\csname eqalign\endcsname\relax
\def\eqalign#1{\null\vcenter{\def\\{\cr}\openup\jot\m@th
  \ialign{\strut$\displaystyle{##}$\hfil&$\displaystyle{{}##}$\hfil
      \crcr#1\crcr}}\,}
\fi
}

\def\BreakURLText#1{\@tfor\brk@tempa:=#1\do{\brk@tempa\hskip0pt}}
\let\lt=<
\let\gt=>
\def\processVert{\ifmmode|\else\textbar\fi}

\@ifundefined{subparagraph}{
\def\subparagraph{\@startsection{paragraph}{5}{2\parindent}{0ex plus 0.1ex minus 0.1ex}%
{0ex}{\normalfont\small\itshape}}%
}{}

\newcommand\role[1]{\unskip}
\newcommand\aucollab[1]{\unskip}
  
\@ifundefined{tsGraphicsScaleX}{\gdef\tsGraphicsScaleX{1}}{}
\@ifundefined{tsGraphicsScaleY}{\gdef\tsGraphicsScaleY{.9}}{}
\def\checkGraphicsWidth{\ifdim\Gin@nat@width>\linewidth
	\tsGraphicsScaleX\linewidth\else\Gin@nat@width\fi}

\def\checkGraphicsHeight{\ifdim\Gin@nat@height>.9\textheight
	\tsGraphicsScaleY\textheight\else\Gin@nat@height\fi}

\def\fixFloatSize#1{}%\@ifundefined{processdelayedfloats}{\setbox0=\hbox{\includegraphics{#1}}\ifnum\wd0<\columnwidth\relax\renewenvironment{figure*}{\begin{figure}}{\end{figure}}\fi}{}}
\let\ts@includegraphics\includegraphics

\def\inlinegraphic[#1]#2{{\edef\@tempa{#1}\edef\baseline@shift{\ifx\@tempa\@empty0\else#1\fi}\edef\tempZ{\the\numexpr(\numexpr(\baseline@shift*\f@size/100))}\protect\raisebox{\tempZ pt}{\ts@includegraphics{#2}}}}

\setkeys{Gin}
{width=\checkGraphicsWidth,height=\checkGraphicsHeight,keepaspectratio}
\DeclareMathAlphabet{\mathpzc}{OT1}{pzc}{m}{it}

\def\URL#1#2{\@ifundefined{href}{#2}{\href{#1}{#2}}}

\def\UrlOrds{\do\*\do\-\do\~\do\'\do\"\do\-}%
\g@addto@macro{\UrlBreaks}{\UrlOrds}

\edef\fntEncoding{\f@encoding}

\makeatother

\newif\ifmultipleabstract\multipleabstractfalse%
\counterwithout{figure}{section}
    \makeatletter
\def\ead{\@ifnextchar[{\@uad}{\@ead}}
\gdef\@ead#1{\bgroup
   \def\_{\string\underscorechar\space}
   \def\{{\string\lbracechar\space}
   \def\textdagger{\string\textdagger\space}
   \def\texttildeapprox{\string\texttildeapprox\space}
   \def~{\hashchar\space}
   \def\}{\string\rbracechar\space}
   \edef\tmp{\the\@eadauthor}
   \immediate\write\@auxout{\string\emailauthor
     {#1}{\expandafter\strip@prefix\meaning\tmp}}
  \egroup
}
\gdef\emailauthor#1#2{\stepcounter{ead}
      \g@addto@macro\@elseads{\raggedright
      \let\corref\@gobble
      \eadsep\texttt{#1} (#2)
      \def\eadsep{\unskip,\space}}
}

\makeatother
  
\let\citep\cite
\let\citet\cite

\begin{document}

\nocite{*}

% \begin{abstract} \bfseries \boldmath
% % Start with one or two sentences of background
% This is a simple template to prepare papers in \LaTeX\ for the \textit{Science}-family journals.
% Abstracts start with one or two sentences of background, which should be
% comprehensible to any scientist.
% % Then summarise the results of your observations, experiments, simulations etc.
% The following text should outline the main results of the research.
% Simple mathematical expressions can be included e.g. $a^2+b^2=c^2$.
% % End with a statement of your main conclusions
% The final sentence of the abstract should state the main conclusions and implications.
% \end{abstract}
\begin{frontmatter}
\title{Dual migration modes of unfaulted disconnections on curved twin boundaries}
\cortext[cor1]{
Corresponding author. Email: hao.lyu@dlmu.edu.cn\\  
\hspace*{1.35em}$^\dagger$These authors contributed equally to this work}
\address[inst1]{College of Transportation Engineering, Dalian Maritime University, Lingshui Road, Dalian, 116026, China}
\address[inst2]{Baotou Aluminum Co., Ltd., Baotou, Inner Mongolia 014040, China}
\author[inst1]{Hongrui He$^{\dagger}$}
\author[inst1]{Hao Lyu$^{\dagger}$\corref{cor1}}
\author[inst2]{Xueting Si}

    \begin{abstract}
Abstract 

Grain boundary migration governs microstructural evolution in crystalline materials, directly influencing mechanical properties such as strength and thermal stability. 
Disconnections, which are line defects formed at grain boundaries in response to local curvature, have been identified as critical carriers of boundary migration. 
Here, we investigate the glide of unfaulted disconnections (UFDs) on a $\Sigma 3$(111) coherent twin boundary in aluminum at elevated temperatures using molecular dynamics simulations combined with the Nudged Elastic Band (NEB) method.
Our results reveal a striking bifurcation in migration behavior depending on the disconnection core structure. 
UFDs with a pure edge Burgers vector migrate via a thermally activated double-kink mechanism, exhibiting a migration velocity that increases monotonically with temperature. 
In contrast, UFDs containing a screw dipole component possess an energy barrier approximately eight times lower, and their core structure undergoes a continuous transformation during glide, giving rise to stochastic, bidirectional motion with no systematic temperature dependence. 
These findings demonstrate that the disconnection core structure fundamentally dictates the migration mode and kinetics of twin boundaries, offering new mechanistic insights into disconnection-mediated grain boundary migration.
\end{abstract}
    
\end{frontmatter}

\section{Introduction}
\label{sec1}
%% Labels are used to cross-reference an item using \ref command.

In nanocrystalline materials, grain boundaries (GBs) are the primary microstructural features that govern their unique mechanical and functional properties. Their influence extends well beyond merely obstructing dislocation motion or acting as dislocation sources \cite{hirth_1972,george_1988}. More critically, GBs facilitate plastic deformation through mechanisms like sliding, migration, and rotation, which directly impact thermodynamic stability and the overall mechanical response \cite{howe_role_2009,gottstein_grain_2009,rohrer_grain_2023}. 
In the classical picture, the migration velocity is expressed as the product of a thermally activated mobility and a capillary driving force set by mean curvature $\it{\boldsymbol{\kappa}}$ and GB energy $\it{\boldsymbol{\gamma}}$. 
Both experiments and simulations have exposed critical limitations of this framework. 
Bhattacharya et al. \cite{bhattacharyaGrainBoundaryVelocity2021a} found no correlation between migration velocity and curvature in polycrystalline Ni. 
Zhang et al. \cite{zhang_grain_2020} showed that GB mobilities in polycrystalline iron are uncorrelated with boundary crystallography and can vary in time for a single boundary.
At the atomistic scale, Holm and Foiles \cite{holm_how_2010} demonstrated that a roughening transition produces a bimodal mobility distribution capable of halting grain growth in pure materials, while Homer et al. \cite{homer_classical_2022} revealed that non-Arrhenius and cryogenic migration emerge naturally when intrinsic energy barriers are small. 
Collectively, these findings underscore the need for a mechanistic description of GB migration that accounts for the atomic-scale defect processes governing boundary motion. 
Disconnections, line defects confined to grain boundaries that carry both dislocation and step character \cite{king_effects_1980,hirth_dislocations_1994,hirth_disconnections_2006,sun_disconnections_2018}, have been identified as the elementary carriers governing GB migration.
Their presence accommodates local boundary curvature, while their nucleation, glide, and annihilation constitute the fundamental mechanisms of boundary motion.
The concept of disconnections is not limited to conventional grain boundaries. 
In particular, twin boundaries, as special types of grain boundaries distinguished by their low interfacial energy and mirror-symmetric crystallographic orientation between adjacent grains \cite{CHRIS19951,ZHU20121}, also undergo migration behaviors mediated by disconnections \cite{hirth_steps_1996,hirth_disconnections_2016}. \par
%Recent experimental and atomistic simulation studies \cite{han_grain-boundary_2018,wang_grain_2014,chen_direct_2024,wang_tracking_2022,fang_atomic-scale_2022,zhu_situ_2019,zhu_atomistic_2022} have shed light on the underlying nature of these GB-mediated processes. Notably, many of these processes proceed in a dislocation-like fashion through the motion of disconnections, special line defects confined to grain boundaries that carry both a crystallographic step and a dislocation character\cite{king_effects_1980,hirth_dislocations_1994,hirth_disconnections_2006,sun_disconnections_2018}. The concept of disconnections is not limited to conventional grain boundaries. In particular, twin boundaries, as special types of grain boundaries distinguished by their low interfacial energy and mirror-symmetric crystallographic orientation between adjacent grains \cite{CHRIS19951,ZHU20121}, also undergo migration behaviors mediated by disconnections \cite{hirth_steps_1996,hirth_disconnections_2016}. 

%Experimental observation of disconnection mediated migration
Recent in situ transmission electron microscopy (TEM) nanomechanical experiments have directly provided  atomic-scale evidence that twin boundary migration in nanotwinned fcc metals proceeds via the glide of twinning disconnections (TD). Li et al. \cite{li_twinning_2011,li_influence_2011} reported that the interaction between a lattice dislocation and the coherent twin boundary (CTB) in nanotwinned Cu initiates a cascade of twinning-dislocation multiplication events, generating stepped disconnection arrays that translate the CTB by several atomic planes. Elevated-temperature in situ scanning transmission electron microscopy (STEM) heating experiments on $\Sigma$3\{112\} incoherent twin boundary (ITB) in Au reveal that thermally activated disconnection motion drives fast boundary migration in defect-free interfaces, whereas grain boundary dislocations strongly pin these disconnections and suppress interface mobility \cite{bi_situ_2024}. Complementing these insights, in situ high-resolution TEM deformation studies of nanotwinned Cu \cite{li_migration_2021} captured the migration kinetics of single- and multi-layer TD facets on CTBs, including single-layer TD glide that produces one-step CTB migration, transient two-layer facets generated by TD–TD interactions, and the absence of three-layer ITB-type accumulation. In situ tensile tests on twin-structured Pt nanocrystals \cite {wang_situ_2018} also captured CTB migration, rapid ITB glide, and full/partial dislocation interactions with twin boundaries, demonstrating that disconnection-mediated boundary motion remains a dominant carrier of plasticity even in ultra-strong, twin-structured, high stacking fault energy nanocrystals. In high-temperature Al, in situ TEM revealed the glide of Shockley disconnections with opposite step signs, whose competing step heights cancel shear-coupled motion and produce GB sliding rather than twin migration \cite{larranaga_high_2023}. \par
%To further elucidate twin-boundary migration, recent atomistic simulations offer a complementary perspective by resolving the core structures of twinning disconnections, quantifying their shear-coupled motion through homogeneous and heterogeneous disconnection dipole glide, and revealing the thermally activated characteristics of disconnection-mediated twin boundary migration at elevated temperatures.
Atomistic simulations complement experimental observations by resolving the core configurations, formation energies, and interaction mechanisms of twin boundary disconnections at the atomic scale.
Atomistic and high-resolution TEM investigations by Marquis and Medlin \cite{marquis_structural_2005} demonstrate that $\Sigma$3 twin disconnections exhibit a structural duality, forming either compact cores or dissociated Shockley partial configurations depending on the Burgers vector orientation and the associated elastic interaction energies.  
Subsequent molecular statics studies \cite{hu_disconnection-mediated_2021,hu_discontinuous_2023} have identified additional twin disconnection core structures and characterized their stress fields, strain distributions, and segregation behavior. Further analysis reveals that elevated temperatures and progressive solute clustering increasingly immobilize these disconnections, thereby suppressing twin boundary migration \cite{hu_stability_2025}.\par
Beyond core structural characterization, atomistic simulations have provided compelling evidence for disconnection mediated twin boundary migration under diverse driving conditions.
By applying a synthetic driving force derived from the energy difference between bicrystals, Deng et al. \cite{deng_size_2017} demonstrated that grain boundary migration is governed by the nucleation and glide of disconnections, and that this mechanism operates not only under shear coupled conditions but also under synthetic normal driving forces, with the resulting boundary mobility exhibiting pronounced size and temperature dependence. Thomas et al. \cite{thomas_coupling_2019} further showed that disconnection nucleation and the subsequent migration of coherent twin boundaries (CTBs) can proceed during annealing of polycrystalline metals, even in the absence of an applied shear stress. Wang et al. \cite{wang_situ_2025} revealed that two non-coplanar Shockley partial dislocations nucleate on opposite sides of a CTB and converge to form a two-layer disconnection step, which subsequently undergoes diffusion assisted plane exchange, enabling pure sliding or one atomic spacing CTB migration. 
Efforts to unify these observations into a comprehensive theoretical framework have made significant progress. Han et al. \cite{han_grain-boundary_2018,thomas_reconciling_2017} derived a generalized disconnection model for grain boundary migration from systematic atomistic simulations, providing a unified kinetic description applicable to diverse boundary geometries and driving conditions. Extending this framework to incorporate multiple coexisting disconnection modes captures the non-monotonic temperature dependence of grain boundary mobility and reveals activation of modes beyond conventional double-kink motion \cite{chenTemperatureDependenceGrain2020a}. Nevertheless, when a boundary is curved or embedded within a polycrystalline network, additional complexities arise that are not fully addressed by current disconnection models \cite{rohrer_grain_2023}.
In particular, it remains unclear how the disconnection core structure and the energy density difference between the adjoining bicrystals jointly control disconnection motion and twin boundary migration in the absence of shear-coupled driving stresses at elevated temperatures.\par
In this work, we first investigate the core structures and energies of various disconnection dipoles using an effective dislocation-dipole description.  
We then equilibrate these systems at elevated temperatures in the absence of shear-coupled stress and evaluate the energy barriers for initiating disconnection motion in dipoles with either screw or pure-edge character. 
For each case, we determine whether the two disconnections migrate toward one another and annihilate, producing a single-layer-height grain boundary migration event, or instead remain separated as a stable dipole. 
Finally, we analyze the conditions governing the distinct migration modes exhibited by different disconnection types and compare the resulting kinetics with existing theoretical, atomistic, and experimental accounts, highlighting the relative roles of atomic shuffling and double-kink mechanisms in disconnection-driven grain boundary migration.
\section{Methods}
\label{sec2}
\subsection{Atomistic simulations}
\label{subsec2.1}
\begin{figure}[h!]%% placement specifier
%% Use \includegraphics command to insert graphic files. Place graphics files in 
    \centering
    \hspace*{-1.5cm}
    \includegraphics[width=1\textwidth]{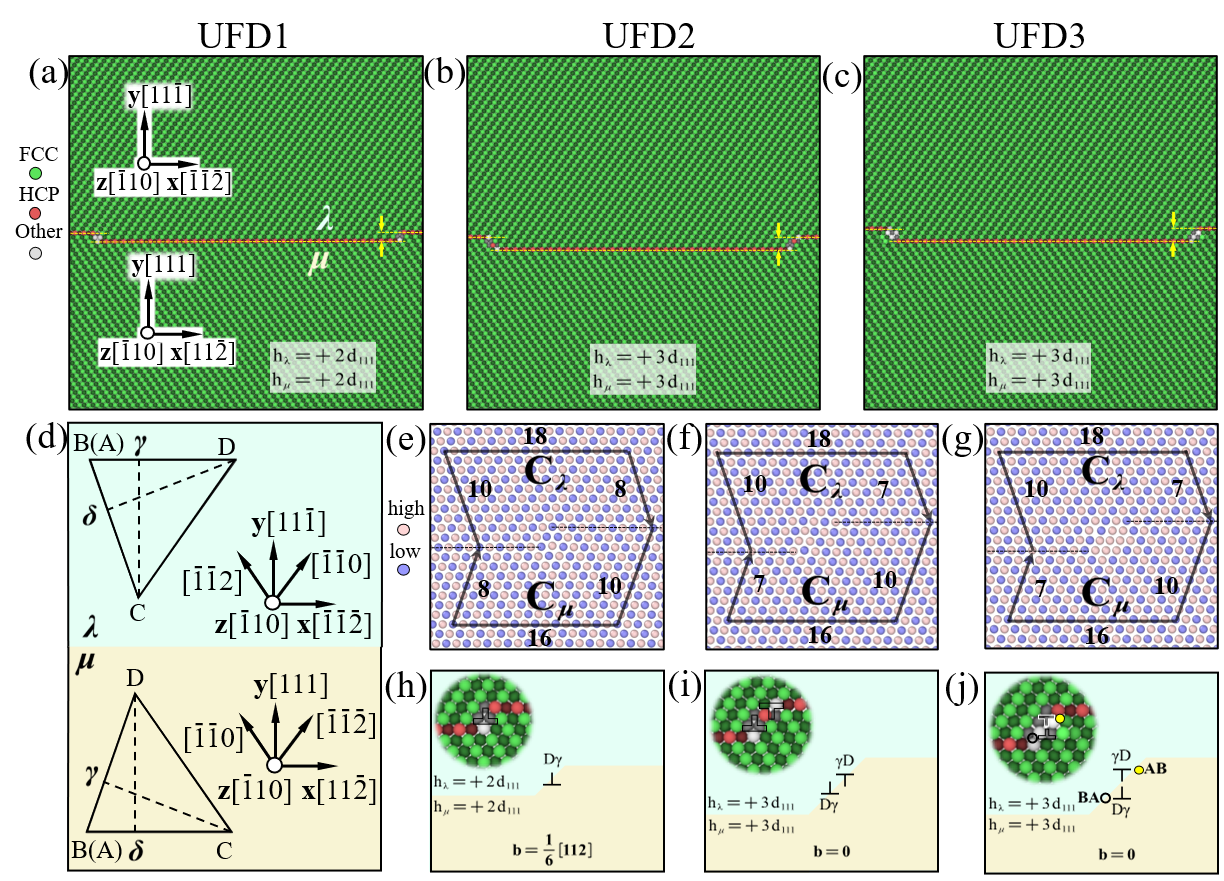}
%% Use \caption command for figure caption and label.
    \caption{Atomistic structures of the $\Sigma 3\,(111)$ coherent twin boundary and associated symmetric unfaulted disconnection (UFD) configurations.
    (a--c) Atomistic configurations of the twin boundary containing (a) UFD1, (b) UFD2, and (c) UFD3 disconnections. Atoms are colored according to the structure (green:FCC; red:HCP; white:Other)
    (d) Schematic illustration of the coordinate systems and the projection of Thompson’s tetrahedra for the upper ($\lambda$) and lower ($\mu$) grains.
    (e--g) Burgers circuit analyses \cite{hirth_steps_1996} of the disconnection cores for (e) UFD1, (f) UFD2, and (g) UFD3.
    (h--j) Schematic representations of the step heights and core structures corresponding to UFD1, UFD2, and UFD3, respectively.
    Defects are identified using adaptive common neighbor analysis (a-CNA)~\cite{stukowskiAutomatedIdentificationIndexing2012}: stacking faults are shown in red, and other defects are shown in gray.
    Atoms are colored by their depth along the viewing direction, with brighter and darker shades indicating shallower and deeper atomic positions, respectively.
}\label{disconnection_core}
%% https://en.wikibooks.org/wiki/LaTeX/Importing_Graphics#Importing_external_graphics
\end{figure}

Atomistic simulations were performed using the Large-scale Atomic/Molecular Massively Parallel Simulator (LAMMPS) code \cite{thompsonLAMMPSFlexibleSimulation2022b} together with an embedded-atom-method (EAM) potential for Al \cite{mishinInteratomicPotentialsMonoatomic1999}. 
An Al bicrystal model containing approximately 489,600 atoms was constructed by joining two crystals whose crystallographic orientations correspond to a CTB (Fig.~\ref{disconnection_core}). 
The two crystals were first generated using Atomsk \cite{hirelAtomskToolManipulating2015}, then cut along identical prescribed curved interfaces and joined together with periodic boundary conditions applied in all directions. 
A rigid-translation method was then used to adjust the relative lateral positions of the two crystals within the x–z plane, followed by energy minimization of the entire simulation cell in which both the box dimensions and atomic positions were relaxed to achieve an overall stress-free state and an equilibrated grain boundary structure. 
Through this procedure, three distinct disconnection configurations with stable core structures were obtained, free of stacking faults emanating from the grain boundary. 
Hu et al.\cite{huDiscontinuousSegregationPatterning2023b} refer to this configuration as the unfaulted disconnection (UFD). Each disconnection dipole comprises a pair of UFDs whose Burgers vectors are equal and opposite, yielding a zero net Burgers vector. Burgers vector circuit analyses \cite{hirth_steps_1996} were applied to characterize the Burgers vector and step height of each constituent UFD. To further elucidate the disconnection core structures, stress analysis was performed to identify the equivalent dislocation associated with each individual disconnection.
%Three pairs of distinct UFD configurations were built, including: (i)fig\ref{disconnection_core}(a) show a sample with one pairs of unfaulted disconnection with a Burgers vector, $b = \frac{1}{6}\left[ 112 \right]$, and have a step with $2d_{111}$; (ii)fig\ref{fig2.1}(b,c) show two samples with one pairs of unfaulted disconnection with a Burgers vector, $b = 0$ and have a step with $3d_{111}$,but these two disconnections possess distinct microstructures.The different structures show some special properties.

Grain boundary activities were then studied at various temperatures maintained by a Nosé–Hoover thermostat \cite{nose2002molecular,hooverCanonicalDynamicsEquilibrium1985,hooverConstantpressureEquationsMotion1986}. 
For each system, five independent simulations were performed with different random seeds for the initial atomic velocity distribution to ensure statistical robustness. To study the effect of energy density differences across the boundary, the efficient calculation of the orientational driving force (ECO) method \cite{schrattEfficientCalculationECO2020,ulomekEnergyConservingOrientational2015} was employed, in which an additional energy of $-\varDelta E/2$ or $\varDelta E/2$ is assigned to atoms in Grain 1 or Grain 2, respectively, according to their crystal orientation, such that the total energy of the system remains approximately unchanged. Prior to applying the ECO driving force, the system was equilibrated at the target temperature using the NVT ensemble. The simulation was then switched to the NPT ensemble to relieve the pressure generated during heating.

The energy barriers for disconnection motion were determined using the nudged elastic band (NEB) method to calculate the minimum energy path (MEP) for grain boundary disconnection migration.
A total of 48 replicas were used to resolve the elementary migration process in which the disconnection advances by either $\frac{1}{4}\left[ 112 \right] $ or $\frac{1}{2}\left[ 112 \right] $ along the grain boundary. Convergence of the energy profile with respect to the number of replicas was verified and is documented in the Supplementary Material. 
Atomistic configurations were visualized using the Open Visualization Tool (OVITO) \cite{stukowskiVisualizationAnalysisAtomistic2010b}, and defects in the bicrystal were identified through adaptive common neighbor analysis \cite{stukowskiAutomatedIdentificationIndexing2012}. The disconnections are approximated as straight line segments aligned with the z-axis. Atoms classified as "other" by the structural analysis are separated into two clusters using k-means clustering\cite{pedregosaScikitlearnMachineLearning}, and the resulting cluster centroids serve as effective descriptors of the disconnection positions. The disconnection pair separation is then calculated from these centroids, enabling evaluation of the relative coupled migration rate between the two disconnections.

\subsection{Dislocation analysis at disconnections}

A Burgers vector circuit analysis \cite{hirth_steps_1996} was employed to characterize the interfacial dislocation by constructing a closed-loop path around each disconnection, as illustrated in Fig.\ref{disconnection_core}(e-g). When this closed path is mapped onto a perfect reference lattice, the resulting closure failure defines a Burgers vector $\bold{b}$ as described
%\begin{figure}[h]%% placement specifier
%% Use \includegraphics command to insert graphic files. Place graphics files in 
   % \centering
    %%\hspace*{-1.5cm}
   % \includegraphics[width=1\textwidth]%{Figures/sigma3_111_burger.png}
%% Use \caption command for figure caption and label.
   % \caption{Atomistic structure of the $\Sigma 3\left( 111 \right)$ twin boundary and corresponding disconnections.}\label{fig2.2}
%\end{figure}
\begin{equation}
\mathbf{b}=-\left( \mathbf{C}_{\mathbf{\lambda }}+\mathbf{PC}_{\mathbf{\mu }} \right) 
\end{equation}
where $\bold{C_\lambda}$ and $\bold{C_\mu}$ are the circuit paths in the two crystals. $\bold{P}$ denotes the relationship between the coordinate frames of the two crystals ($\mu$ to $\lambda$). Here, our chosen coordinate system $\bold{P}$ is given by:
\begin{equation}
    \mathbf{P}={\frac{1}{3}}\left[ \begin{matrix}
	2 &		-1 &		-2\\
	-1 &		2 &		-2\\
	-2 &		-2 &		1\\
\end{matrix} \right]
\end{equation}

\begin{figure}[h!]%% placement specifier
%% Use \includegraphics command to insert graphic files. Place graphics files in 
    \centering
    %%\hspace*{-1.5cm}
    \includegraphics[width=1\textwidth]{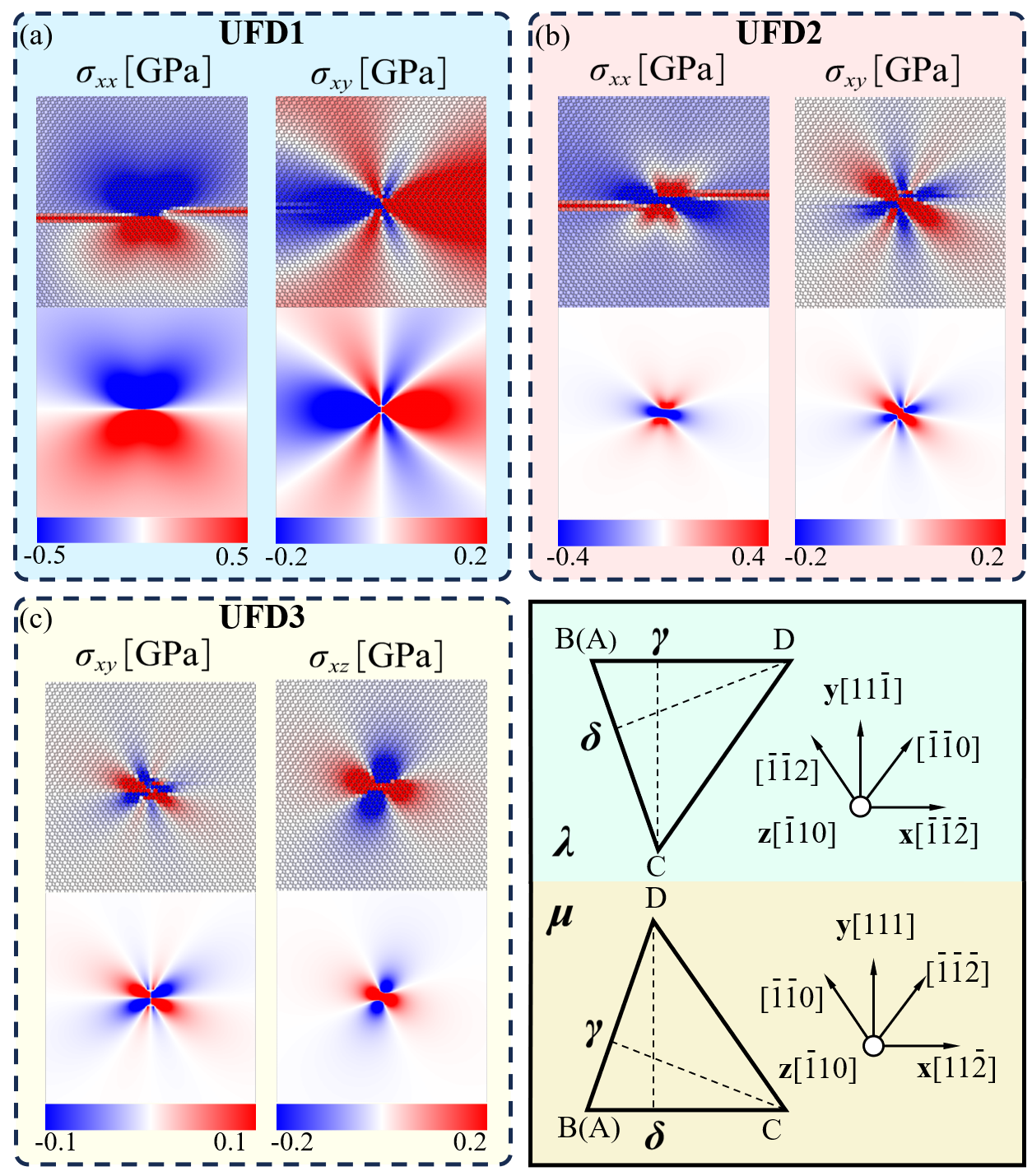}
%% Use \caption command for figure caption and label.
    \caption{Comparison between equivalent dislocation stress fields calculated from continuum elasticity solutions and atomistic stress fields obtained from simulations for (a) UFD1-, (b) UFD2-, and (c) UFD3-type disconnections.}\label{disconnection_strain}
\end{figure}

Based on the dislocation character of the disconnections, stress analysis of the equivalent dislocations was performed using the point stress calculation implemented in a discrete dislocation dynamics framework \cite{COTTRELL1953205}, as illustrated in Fig.\ref{disconnection_strain}.

% The interaction energy of two disconnections are calculated based on the interaction energy of two parallel dislocation using
% These self energy of disconnections compute with the formula
% \begin{equation}
% \label{eq:W_core}
% \begin{split}
% W_{\text{self}} &= \frac{\mu \pi}{8} \left[
% \left( \frac{2\| \mathbf{b}_{e} \|^2 LN}{1-\nu} \right)
% + \| \mathbf{b}_{s} \|^2 \left( 2LN - \frac{3-\nu}{1-\nu} \right) (L_a - a)
% \right] \\
% &\quad + \frac{\mu}{4\pi} \left(
% \frac{\| \mathbf{b}_{e} \|^2}{1-\nu}
% + \| \mathbf{b}_{s} \|^2
% \right) L
% \ln\left( \frac{L}{ea} \right)
% \end{split}
% \end{equation}

% $L_a=\sqrt{L^2+a^2},N=\ln \left( \frac{L_a+L}{a} \right) ,a=\frac{\left\| \boldsymbol{b} \right\|}{0.2}\cdot 1.17 $
% %三种(四种)阶错应力场以及使用MATLAB软件绘制的等效位错应力场如图
% \begin{equation}
% \label{eq:W_interaction}
% \begin{split}
% W_{interaction} &= \left[ -\frac{\mu}{2\pi}\left( \mathbf{b}_{\mathbf{1}}\times \mathbf{b}_{\mathbf{2}} \right) \cdot \left( \boldsymbol{\xi }_{\boldsymbol{1}}\times \boldsymbol{\xi }_{\boldsymbol{2}} \right) +\frac{\mu}{4\pi}\left( \mathbf{b}_{\mathbf{1}}\cdot \boldsymbol{\xi }_{\boldsymbol{1}} \right) \left( \mathbf{b}_{\mathbf{2}}\cdot \boldsymbol{\xi }_{\boldsymbol{2}} \right) \right] \int\limits_{x_1}^{x_2}{dx}\int\limits_{y_1}^{y_2}{\frac{dy}{R}}
% \\
% &\quad+\frac{\mu}{4\pi \left( 1-\nu \right)}\left( \mathbf{b}_{\mathbf{1}}\times \boldsymbol{\xi }_{\boldsymbol{1}} \right) \cdot \mathbf{T}\cdot \left( \mathbf{b}_2\times \boldsymbol{\xi }_2 \right) 
% \end{split}
% \end{equation}

\section{Results and discussion}
\label{sec3}
\subsection{Symmetric unfaulted disconnections on curved twin boundary}
Al bicrystal atomic models containing a curved twin boundary were constructed following the methodology described in the Methods section ~\ref{sec2}. After rigid-body translation of the grain boundary plane and energy minimization, multiple equilibrated twin boundary structures were obtained (see Supplementary Material); among these, only configurations featuring symmetric unfaulted disconnections (UFDs) are presented in Fig.\ref{disconnection_core}. Burgers circuit analysis was first employed to characterize the interfacial defect content of these configurations. The UFD1 structure contains a disconnection dipole in which each constituent disconnection has pure edge character. In UFD2, each disconnection within the dipole is itself described by an edge dislocation dipole rather than a single dislocation. The UFD3 configuration exhibits a more complex disconnection structure, in which each disconnection comprises a coupled edge and screw dislocation dipole. 
To further corroborate the identified dislocation characters, the stress fields associated with the different UFD configurations were quantitatively analyzed using elastic solutions for finite dislocation segments. The shear modulus (27 GPa) and Poisson's ratio (0.325) used in these calculations were extracted from the atomistic model to ensure consistency between the continuum and atomistic descriptions. Overall, continuum elastic theory successfully reproduces the characteristic stress fields of the disconnections, with minor deviations attributable to the presence of the twin boundary.\par
Consistent with these atomistic observations, disconnections analogous to those in the UFD1 and UFD2 configurations have been resolved by high-resolution transmission electron microscopy (HRTEM) in nanotwinned Cu, where their migration was directly captured \cite{li_migration_2021}. Similar UFD1- and UFD2-type disconnection configurations have also been directly observed at coherent twin boundaries (CTBs) in Pt \cite{wang_situ_2025}. In addition, Zhu et al. \cite{zhu_atomistic_2022} reported stress-induced lateral motion of UFD1 disconnections along with multi-layer disconnections in Au. While direct experimental evidence of UFD3-type disconnections with mixed dislocation character remains unavailable, related mixed dislocations with zero net Burgers vector involving screw dislocation dipoles have been observed during detwinning and twin growth at incoherent twin boundaries (ITBs) in Cu \cite{wang_detwinning_2010}. Furthermore, disconnections and grain boundary structures containing screw components have been observed in Au \cite{chenRevealingGrainBoundary2024a}, supporting the plausibility of such mixed-character disconnection configurations.

\subsection{The migration of curved twin boundary at elevated temperatures}

\begin{figure}[hbt!]%% placement specifier
%% Use \includegraphics command to insert graphic files. Place graphics files in 
    \centering
    %%\hspace*{-1.5cm}
    \includegraphics[width=1\textwidth]{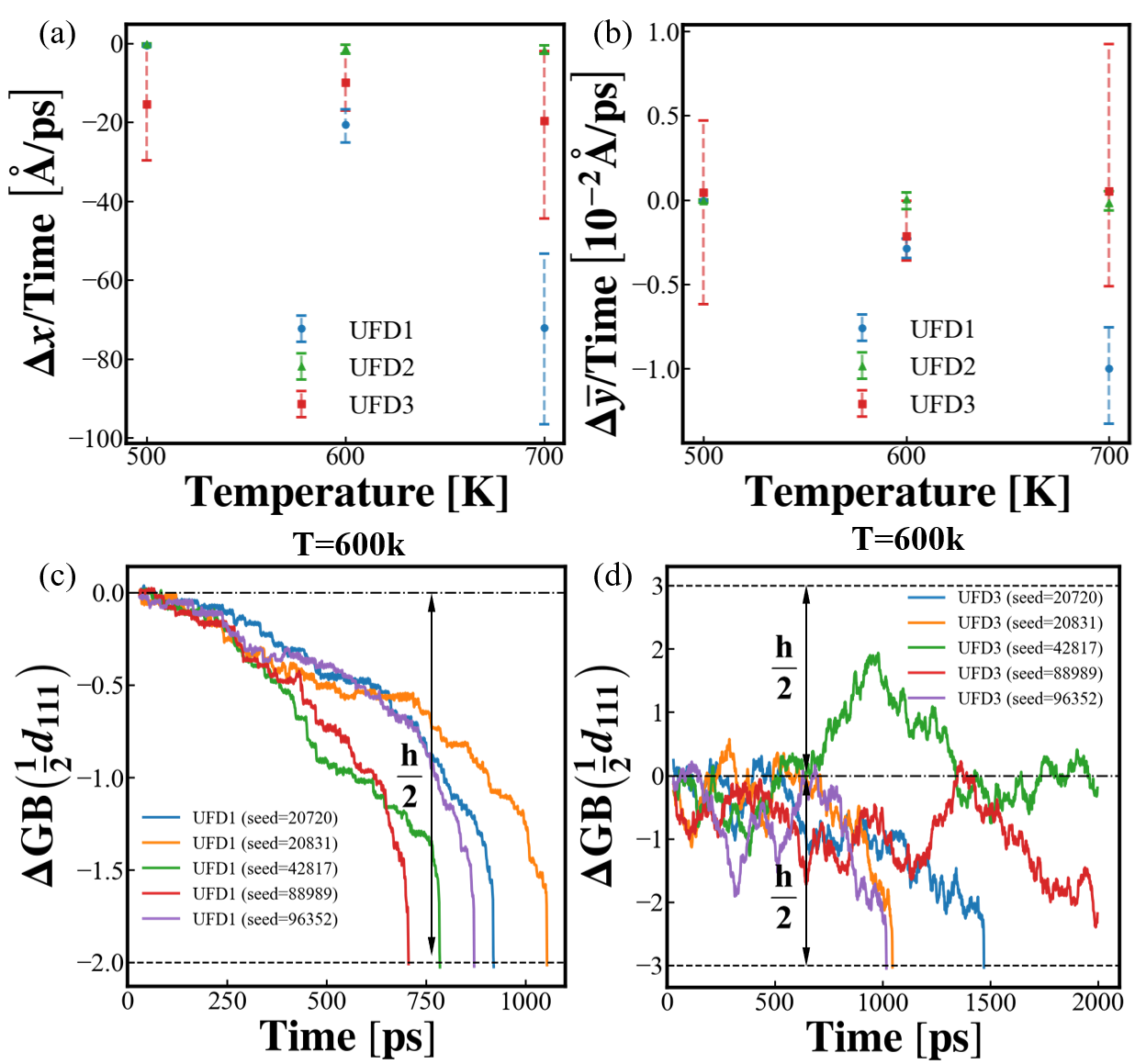}
%% Use \caption command for figure caption and label.
    %\caption{Displacement of the UFD1 grain boundary flattening time on random seeds at 600 K.}\label{migration_rate}
    \caption{(a) Temperature dependence of the mean disconnection velocities ($\Delta x/$Time) for the UFD1, UFD2, and UFD3 configurations, calculated as the relative approach speed of the two disconnections along the grain boundary plane. (b) Corresponding mean grain boundary migration rate ($\Delta \bar{y}/$Time) as a function of temperature, determined from the step height of the disconnections divided by the closing time. (c,d) Time evolution of the grain boundary displacement at 600 K for the (c) UFD1 and (d) UFD3 configurations, obtained from five independent simulations with different random velocity seeds. The trajectories highlight the distinct migration behaviors and the tendency of the grain boundary to evolve toward a flattened configuration. Reference lines denote the initial twin boundary position (dash-dotted) and the final twin boundary position after disconnection annihilation (dashed)}\label{migration_rate}
\end{figure}

Twin boundary migration in all three configurations was systematically investigated over a range of temperatures. To quantify the migration behavior, the migration rates of both the disconnections and the twin boundary were determined by tracking the relative separation between two disconnections and the evolution of the average position of all twin-boundary atoms along the $\langle111\rangle$ direction (Fig.~\ref{migration_rate}a, b). Across the entire temperature range investigated, the UFD2 configuration exhibits no measurable grain boundary migration or disconnection motion, in contrast to the UFD1 and UFD3 configurations. This immobility is attributed to its locally zero net Burgers vector and low disconnection line energy (see Supplementary Fig. 2).
In contrast, the UFD1 configuration exhibits the onset of migration above 600 K, whereas grain boundary migration in the UFD3 configuration initiates at a lower temperature of approximately 500 K. Further analysis reveals that both the grain boundary migration rate and the glide speed of edge disconnections increase monotonically with temperature, while no systematic temperature dependence is observed for the mixed-character disconnections. 
The time evolution of the mean twin boundary position (Fig.~\ref{migration_rate}c) shows that the UFD1 configuration undergoes clear directional migration toward the bottom grain. As the disconnections migrate, the twin boundary advances by $h/2$, during which the separation between the two disconnections decreases correspondingly until they eventually annihilate. 
The migration rate increases sharply as the two disconnections approach each other, leading to rapid completion of grain boundary migration. Fig.~\ref{migration_rate}d shows fluctuations in the mean grain boundary position, reflecting the stochastic back-and-forth motion of the disconnections. Additional simulations reveal that annihilation of the disconnection dipole can occur on either side, indicating that the grain boundary is capable of migrating both upward and downward. These results suggest that disconnection motion on the CTB is governed by multiple mechanisms that depend on the character of the disconnections.
When the net Burgers vector vanishes locally within the disconnection structure, the interaction energy between disconnections no longer plays a dominant role, particularly during the late stage of migration. Moreover, the presence of a screw component introduces additional modes of motion that differ fundamentally from the double-kink mechanism proposed for pure edge disconnections in previous studies \cite{hanGrainboundaryKineticsUnified2018,chenTemperatureDependenceGrain2020a}.

\subsection{ The motion of UFD1-type disconnection} \label{sec:UFD1}
\begin{figure}[hbt!]%% placement specifier
    \centering
    %%\hspace*{-1.5cm}
    \includegraphics[width=0.96\textwidth]{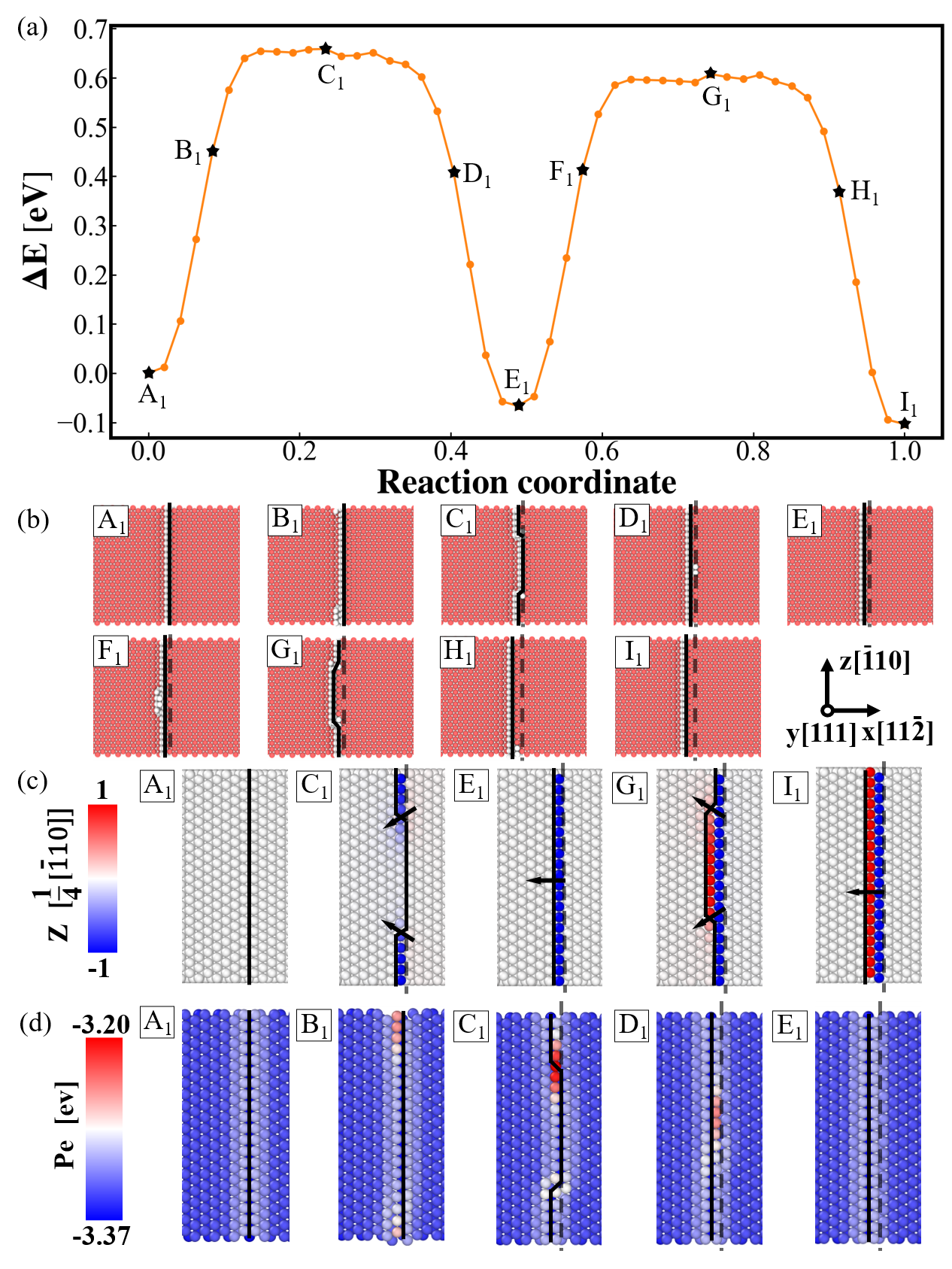}
%% Use \caption command for figure caption and label.
    \caption{NEB calculations for UFD1-type grain boundary (GB) migration. (a) Minimum energy path (MEP) for UFD1-type GB migration over a distance of $\tfrac{1}{2}[112]$ at 600 K. (b) Top-view atomistic configurations of the UFD1-type disconnection along the MEP. (c) Displacement field of atoms along the $z$ direction, from $\mathbf{A}_1$ to $\mathbf{I}_1$.(d) %Averaged potential energy of the double-kink configuration,% calculated within a cutoff radius of 5? Å, 
    Per-atom potential energy of the double-kink configuration,
    from $\mathbf{A}_1$ to $\mathbf{E}_1$}\label{UFD1 NEB L move2}
\end{figure}

NEB simulations were performed to investigate the motion of the left disconnection of the dipole, using energy-minimized configurations before and after one step of disconnection motion at 600 K as the initial and final states. The calculated energy profile (Fig.~\ref{UFD1 NEB L move2}a) reveals an energy barrier of approximately 0.65 eV. The motion proceeds via a double-kink mechanism \cite{han_grain-boundary_2018}. As shown in Fig.~\ref{UFD1 NEB L move2}b and c, a kink pair first nucleates at B$_1$, preferentially in a region of relatively high local potential energy. Subsequent kink propagation drives the entire atoms along the disconnection to advance along the $z$ direction by $\tfrac{1}{4}[101]$,which in turn results in the motion of the disconnection by $\tfrac{1}{4}[112]$ along the x direction, as illustrated by the evolution from C$_1$ to E$_1$.
The two kinks exhibit distinct energetic characteristics. One is associated with a locally compact atomic configuration and a markedly elevated potential energy, whereas the other adopts a less compact structure with lower potential energy. Upon kink annihilation, the local energy distribution relaxes to a state energetically comparable to the initial configuration. Repetition of this process produces an additional displacement of $\tfrac{1}{4}[011]$ of the atomic column, resulting in a net disconnection motion of $\tfrac{1}{2}[112]$.

At elevated temperatures, disconnection dipole motion becomes increasingly governed by thermally activated kink nucleation and propagation. The corresponding kink pair nucleation rate is given by Eq.~\ref{eq:thermalkink}.
\begin{equation}
\label{eq:thermalkink}
c_k^+c_k^-=\frac{1}{a^2}\exp \left( -\frac{2F_k}{kT} \right)
\end{equation}
where the Helmholtz free energy $2F_k$ of a kink pair consists of an energetic contribution $2U_k$ and an entropic contribution $2S_k$. The energetic term $2U_k$ comprises elastic and core misfit energies associated with kink formation, kink interactions, twin boundary energy, and interaction between disconnection dipoles. Based on the NEB calculations, $2U_k = 0.45$ eV is adopted as a representative kink-pair nucleation energy barrier. Using this value, Supplementary Fig.~2 shows that kink pair nucleation is essentially absent at 500 K, indicating that the initiation of disconnection motion is highly unlikely at this temperature. With increasing temperature, thermal fluctuations progressively reduce the effective free energy barrier for kink-pair nucleation. Accordingly, the free energy can be expressed as
\begin{equation}
\label{eq:FEkink}
2F_k=2U_k-2TS_k
\end{equation}
The energetic term $U_k$ is independent of temperature, whereas the entropy term can be approximated as $S_k \cong k \ln \left( \tfrac{kT}{\hbar \omega } \right)$ . This entropy-driven reduction in $2F_k$ leads to an exponential increase in the kink-pair nucleation rate, thereby promoting localized kink-pair formation and facilitating disconnection motion. Such thermally assisted, local nucleation events are consistent with the behavior observed at the onset of disconnection motion or grain boundary migration observed in Fig.~\ref{migration_rate}. 

\begin{figure}[hbt!]%% placement specifier
    \centering
    %%\hspace*{-1.5cm}
    \includegraphics[width=1.0\textwidth]{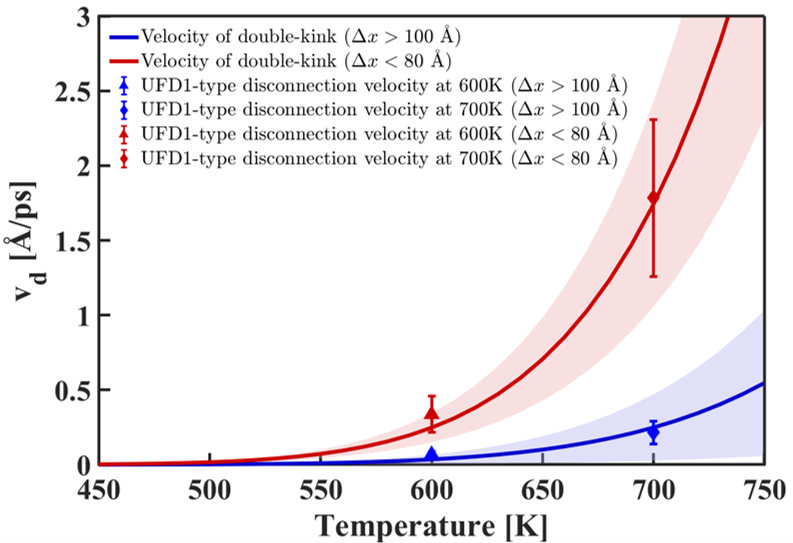}
%% Use \caption command for figure caption and label.
    \caption{Temperature dependence of UFD1-type disconnection velocity for two separation regimes: far-field ($\Delta x>100\mathring{A}$, blue) and near-field ($\Delta x<80\mathring{A}$, red). Symbols with error bars represent the mean velocities averaged over five independent simulations at 600 K and 700 K, while solid curves correspond to the predictions of the double-kink nucleation model. Shaded regions indicate the uncertainty in theoretical predictions, reflecting the variation of interaction stress over the dipole separation ranges in the far-field and near-field regimes}\label{UFD1 velocity}
\end{figure}

In our atomistic simulations, the disconnection velocity is calculated directly from the trajectory as
\begin{equation}
\label{eq:v_def}
\Bar{v}=\frac{2\min \left( x_r-x_l,L_x+x_l-x_r \right)}{\varDelta tL_x}
\end{equation}
where $\min \left( x_r-x_l,L_x+x_l-x_r \right)$ denotes the displacement of the tracked disconnections on the glide direction, determined from the time evolution of the separation between the left ($x_l$) and right ($x_r$) disconnections of the dipole, $\Delta t$ is the corresponding time interval. Because the interaction energy between the two disconnections increases as they approach each other, the motion can be divided into two distinct stages. In the first stage, the interaction energy is negligible over the considered $\Delta t$, with the disconnection spacing ranging from the initial separation down to approximately $10\,\mathrm{nm}$. In the second stage, as the disconnections approach more closely ($\min \left( x_r-x_l,L_x+x_l-x_r \right) \approx 8\,\mathrm{nm}$ until annihilation), the interaction energy becomes significant. 
For the thermally activated double-kink mechanism, the corresponding average disconnection velocity can be expressed as
\begin{equation}
\label{eq:v_kink}
\Bar{v}=\frac{ h^2 L}{a^2 k T} \left[ (\sigma +\sigma_{int} )b+\psi h\right] D_k
\exp\left(-\frac{2U_k-2TS_k}{kT}\right),
\end{equation}
where the interaction energy between the disconnections is incorporated into the effective driving stress $\sigma$ acting on the disconnection during the second stage. Accordingly, as shown in Fig.~\ref{UFD1 velocity}, the applied stress $\sigma$ is set as 0, two different values of $\sigma_{int}$ are adopted for the two stages of motion based on the disconnection separation. Here, $b$ is the Burgers vector, $h$ is the kink-pair height, $L$ is the disconnection length, $a$ is the lattice parameter, $k$ is the Boltzmann constant, and $T$ is the temperature. The kink diffusivity $D_k$ is given by

\begin{equation}
\label{eq:D_k}
D_k = a^2 \nu_D
\exp\left(-\frac{W_m}{kT}\right),
\end{equation}
where $\nu_D$ is the Debye frequency, related to the Debye temperature and taken as $56\,\mathrm{ps}^{-1}$, and $W_m$ is the kink migration energy, which is taken as $0.2\,\mathrm{eV}$ based on the NEB calculations.
As shown in Fig. ~\ref{migration_rate}, the velocity of the disconnection at elevated temperatures can be captured by the double-kink mechanism.

\subsection{The motion of UFD3-type disconnection}\label{sec:UFD3}
\begin{figure}[hbt!]%% placement specifier
%% Use \includegraphics command to insert graphic files. Place graphics files in 
    \centering
    %%\hspace*{-1.5cm}
    \includegraphics[width=1\textwidth]{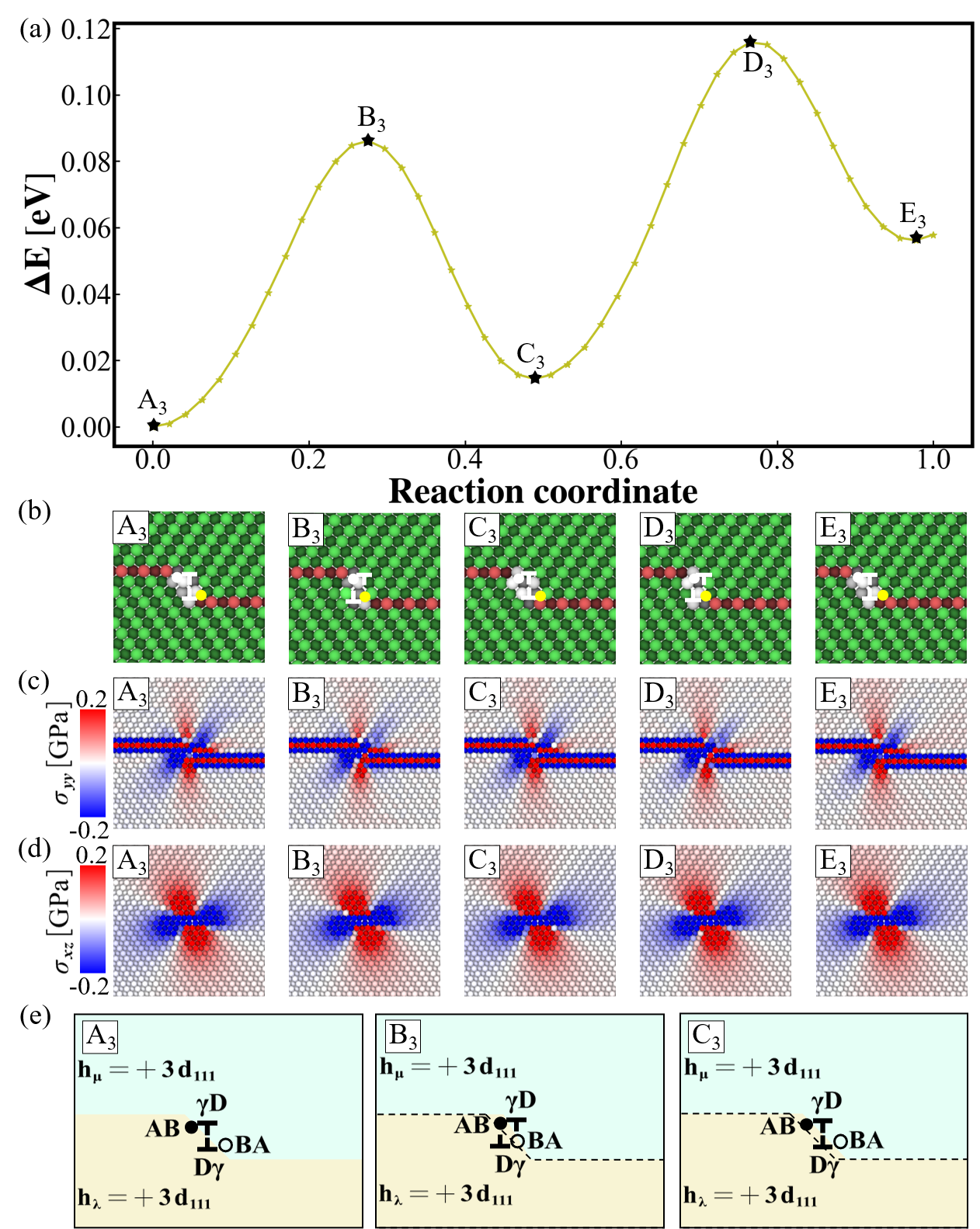}
%% Use \caption command for figure caption and label.
        \caption{NEB calculations for UFD3-type grain boundary (GB) migration. (a) Minimum energy path (MEP) for UFD3-type GB migration over a distance of $\tfrac{1}{2}[112]$ at 600 K. (b) Front-view atomistic configurations of the UFD3-type disconnection along the MEP. (c) Stress field $\boldsymbol{\sigma }_{\boldsymbol{yy}}$ of atoms from $\mathbf{A}_3$ to $\mathbf{E}_3$.(d) Stress field $\boldsymbol{\sigma }_{\boldsymbol{xz}}$ of atoms from $\mathbf{A}_3$ to $\mathbf{E}_3$.(e) Schematic illustration of the disconnection migration sequence from $\mathbf{A}_3$ to $\mathbf{E}_3$}\label{UFD3 NEB}
\end{figure}

The motion of UFD3-type disconnections is highly stochastic, resulting in pronounced positional fluctuations. This is evidenced by the evolution of the mean twin boundary position, which exhibits bidirectional variations throughout the simulation. To further elucidate the underlying mechanism, NEB simulations were performed using energy-minimized configurations before and after disconnection motion, with the disconnection initially located on the left of the dipole. The resulting energy profile for disconnections with screw dislocation character is shown in Fig.~\ref{UFD3 NEB}. The energy barrier for UFD3-type disconnection motion is approximately 0.08 eV, roughly eight times lower than that of UFD1-type disconnections, indicating that UFD3-type disconnection motion is considerably more energetically favorable.
As shown in Fig.~\ref{UFD3 NEB}b, the UFD3 disconnection can be decomposed into a superposition of an edge dislocation dipole and a screw dislocation dipole. Disconnection motion is accompanied by a core structure rearrangement initiated by the displacement of the lower edge dislocation $\boldsymbol{D\gamma}$
together with the associated screw dislocation dipole $\boldsymbol{BA}$ by $\tfrac{1}{4}[101]$, as illustrated in Fig.~\ref{UFD3 NEB}B$_3$.
This is followed by an additional shift of the lower dipoles by $\tfrac{1}{4}[011]$, as shown in Fig.~\ref{UFD3 NEB}C$_3$.
The corresponding evolution of the stress field provides further support for this mechanism. Upon dipole translation, the positive $\sigma_{yy}$ stress field associated with the edge dislocation component shifts laterally, consistent with the observed displacement of the edge dipole, whereas the $\sigma_{xz}$ stress field associated with the screw component remains essentially unchanged. 
This indicates that the screw component does not directly drive the motion but instead provides a stable and compatible core configuration that facilitates lateral migration of the edge dislocation dipole. 
Meanwhile, the metastable core structure at C$_3$ can either overcome the energy barrier to return to the initial configuration A$_3$, or surmount a slightly higher barrier of approximately 0.02 eV (see supplementary Fig 9)to reach E$_3$, which is structurally identical to A$_3$. This reversible core transformation gives rise to the positional fluctuations of the disconnection.\par
Consequently, at elevated temperatures, UFD3-type disconnections can be readily activated through core structure transformations. However, the metastable intermediate state can either revert to the original configuration, returning the disconnection to its initial position, or advance to a structurally identical configuration, completing a forward step. This inherent reversibility of the core transformation underlies the stochastic behavior observed in the simulations, where the mean twin boundary position fluctuates rather than exhibiting sustained directional migration. In contrast to the UFD1 case, no systematic dependence of the migration velocity on temperature is observed, suggesting non-Arrhenius grain boundary migration behavior \cite{homerClassicalEquationThat2022,rheinheimerNonArrheniusBehaviorGrain2015}. 
These positional fluctuations also result in lower net disconnection and grain boundary migration velocities compared with the UFD1 case.
The reduced overall mobility of disconnections with screw character relative to edge-character disconnections is consistent with the observations of Chen et al. \cite{chenRevealingGrainBoundary2024a}, who reported from both in situ TEM experiments and atomistic simulations that edge segments of curved disconnections on the $\Sigma$11 $[110]$ (113) symmetric tilt grain boundary in FCC Au migrate significantly faster than screw segments. They attributed this mobility difference to the higher glide barrier of screw disconnections and the additional surface drag arising from surface step creation during screw segment motion on three-dimensional curved grain boundaries. The physical origin of the reduced mobility, however, differs across systems. Gong et al. \cite{gong_symmetric_2022} performed NEB calculations on $\{10\bar{1}1\}$ coherent twin boundaries in HCP Mg and found that the intrinsic energy barrier for screw twinning disconnections (~0.04 eV) is substantially lower than that for edge twinning disconnections (~0.35 eV), indicating that screw disconnections are more readily activated at the single-disconnection level. This is consistent with our finding that UFD3-type disconnections with screw dislocation character exhibit a much lower energy barrier (~0.08 eV) than UFD1-type edge disconnections (~0.65 eV) on the $\Sigma$3 coherent twin boundary in FCC Al. In the present work, the lower net migration velocity of UFD3-type disconnections, despite their lower activation barrier, originates from the inherent reversibility of the core structure transformation, which produces stochastic positional fluctuations rather than sustained directional migration. Although the crystal structures (FCC Al, FCC Au, HCP Mg), boundary types, and simulation conditions differ across these studies, the results imply that the macroscopic migration rate of a disconnection is not determined solely by its activation energy, but is also influenced by the nature of the migration mechanism and the boundary geometry.
Furthermore, Chen et al. \cite{chenTemperatureDependenceGrain2020a} studied the migration of curved grain boundaries with varying Burgers vectors and step heights, and showed that for boundaries with larger curvature, the migration behavior deviates from the double-kink mechanism, giving rise to non-Arrhenius kinetics. While their work provides a plausible explanation for non-Arrhenius migration at the mesoscale, the analysis does not resolve the local disconnection structure in detail. For grain boundaries with larger curvature, the boundary may accommodate the curvature through combinations of different disconnection structures, such as multiple UFD1, UFD2, or UFD3 configurations. Since each type exhibits distinct migration kinetics, as demonstrated in the present study, a spectrum of migration behaviors would be expected depending on the local disconnection composition. This highlights the need for further investigation of coupled multi-disconnection dynamics in future work.

\subsection{The effect of energy density difference}
Regarding the migration direction of the grain boundary, it is well established that the energy density difference across the boundary, denoted as 
$\psi$, serves as an important driving force for grain boundary migration. In the present atomistic simulations, the lower twin boundary is positioned near the center of the simulation cell, which results in a comparatively larger bottom grain, leaving a higher energy density difference between the upper and lower grain. This energy density imbalance generates a driving force that promotes lateral glide of the disconnections along the boundary plane, which in turn leads to upward migration of the twin boundary. To further examine the role of this driving force, NEB simulations were performed with the initial and final configurations reversed, such that both UFD1- and UFD3-type disconnections migrate toward the center of the simulation cell.
\begin{figure}[hbt!]%% placement specifier
%% Use \includegraphics command to insert graphic files. Place graphics files in 
    \centering
    %%\hspace*{-1.5cm}
    \includegraphics[width=1\textwidth]{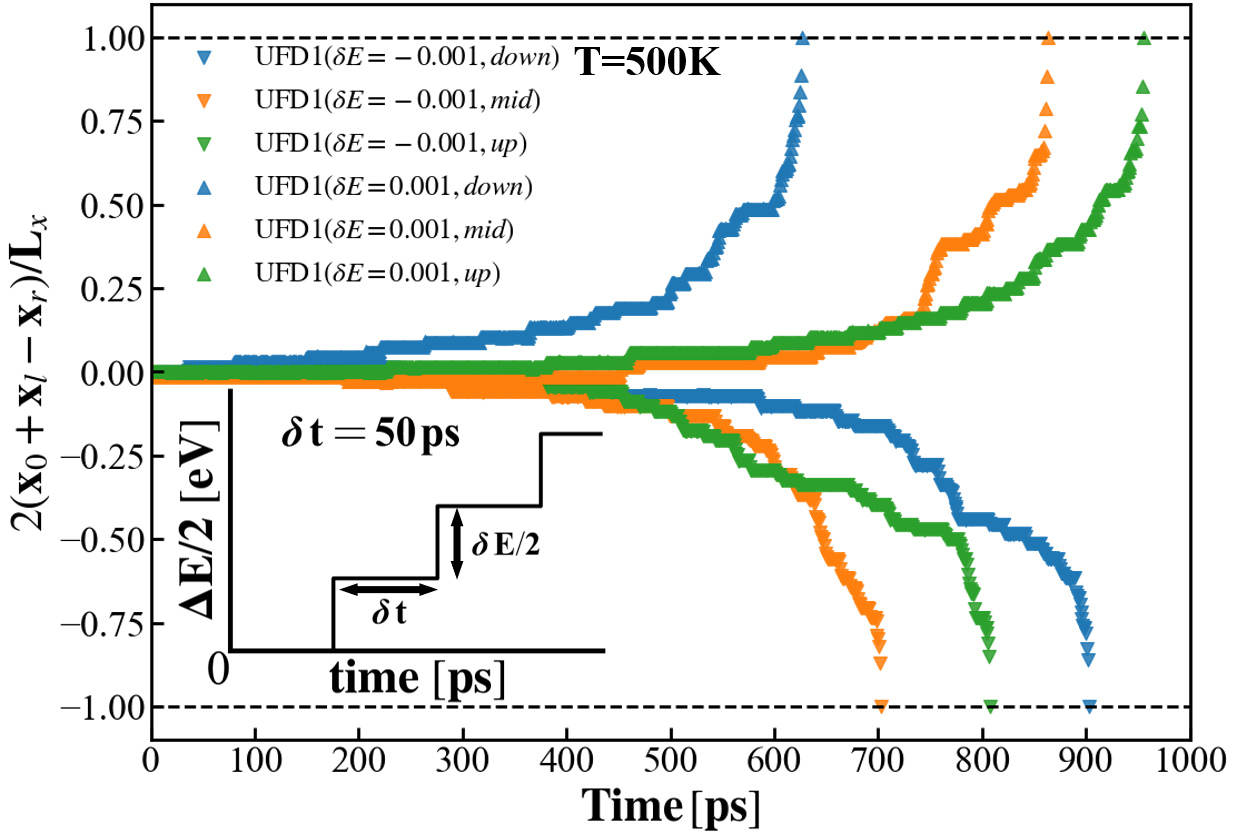}
%% Use \caption command for figure caption and label.
    \caption{Normalized disconnection dipole separation as a function of time for three UFD1 samples with different twin boundary positions (labeled up, mid, and down), corresponding to different grain-size asymmetries, under incrementally applied ECO driving forces at 500 K. Both positive ($\delta$E=0.001eV) and negative ($\delta$E=-0.001eV) energy density differences were applied, yielding six curves. The inset illustrates the stepwise loading protocal with increment $\delta$E/2 applied everty $\delta$t=50 ps}\label{ECO_500k}
\end{figure}

The results demonstrate that disconnection motion against the driving force is associated with a substantially higher energy barrier, making such migration energetically unfavorable(see Supplementary Figs. 6 and 7). As a result, in all UFD1-type and most of UFD3-type configurations, disconnection dipoles preferentially annihilate at the lateral sides of the boundary rather than migrating toward and annihilating at the center. Several methods were examined to quantify the energy density difference across the twin boundary. Nevertheless, no direct and consistent evidence of such an energy density difference could be identified across the grain boundary or in the vicinity of the disconnections at elevated temperatures. Accordingly, to explicitly probe the influence of an energy density difference on disconnection dynamics, a synthetic driving force was introduced to impose a prescribed energy-density gradient across the disconnection. To eliminate thermal effects, all simulations were first conducted at 10 K. As shown in Supplementary Fig. 4, the UFD3-type disconnection migrates in a manner that reduces the imposed energy density, exhibiting lateral motion under a positive energy-density difference and centerward motion under a negative one. In contrast, the UFD1-type disconnection requires a substantially longer incubation time before initiating motion, indicating that a larger driving force is necessary to activate its migration. This behavior is consistent with the higher activation barriers obtained from the NEB simulations discussed in the previous sections. 
Supplementary Fig. 4 indicates that disconnection motion is preferentially activated under a negative energy bias, leading to lateral migration, whereas no detectable centerward motion is observed when the energy bias is reversed. Instead, a single-layer disconnection nucleates at the right-hand disconnection and propagates under an applied synthetic driving force of 0.05 eV. This process leads to the formation of a non-equilibrated disconnection dipole, for which the net Burgers vector is not necessarily zero. This nonzero Burgers vector is evidenced by the evolution of the normal stress component $\sigma_{xx}$ throughout the process. Upon further increasing the driving force to 0.1 eV, the disconnections migrate toward the center of the simulation cell and ultimately undergo annihilation. This indicates that at low temperatures, an abrupt increase in the driving force does not necessarily reverse the migration direction. Instead, it may trigger the nucleation of new disconnection structures and subsequent core structure transformations, leading to non-equilibrium annihilation of disconnection dipoles.

To verify that grain-size asymmetry can produce a finite energy density difference $\psi$ between the two grains, two additional samples with pronounced grain-size differences were constructed (see Supplementary Note 5 for details). The ECO method was applied in both directions at 500 K for all three samples. As shown in Fig.~\ref{ECO_500k}, grain-size asymmetry introduces a directional bias in grain boundary migration by altering the critical driving force required for motion in each direction.
When a positive energy density difference $+\Delta$$E/2$ is applied, the disconnection velocity toward the lateral direction increases, whereas reversing the applied energy density difference leads to a pronounced decrease in disconnection velocity. This asymmetric response indicates the presence of an intrinsic energy density difference $\psi$ arising from grain-size asymmetry.

\section{Conclusion}

In this work, we systematically investigated the migration of curved twin boundaries mediated by disconnections with different character.
Our results demonstrate that curved twin boundary migration can be activated even when the net Burgers vector of the disconnection configuration vanishes, highlighting the critical role of local disconnection dynamics over global Burgers vector balance. Despite sharing identical geometric constraints, different disconnection types exhibit markedly distinct migration behaviors. Disconnections with pure edge character migrate via a thermally activated double-kink mechanism, whereas disconnections with screw character migrate, exhibiting no systematic temperature dependence.

NEB calculations further reveal pronounced differences in activation energies between these migration modes, with pure edge disconnections requiring substantially higher energy barriers than screw disconnections. The lower barrier for screw disconnections, however, does not translate into faster net migration. Instead, the reversibility of the core structure transformation produces highly stochastic motion in which the disconnection may advance or retreat with comparable likelihood, ultimately prolonging the annihilation time of disconnection dipoles.

In addition, the influence of energy density differences on disconnection motion was systematically examined to elucidate the preferential annihilation position of disconnection dipoles, which determines the direction of grain boundary migration.. The motion of both UFD1 and UFD3 disconnections is driven by the energy density difference between the adjoining grains, with UFD1 edge disconnections exhibiting a stronger correlation owing to their thermally activated double-kink mechanism. By tuning the energy density difference, the annihilation site of the disconnection dipole can be shifted laterally, thereby controlling whether the grain boundary migrates toward the upper or lower grain. Furthermore, our simulations reveal that different disconnection structures can transform into one another under certain conditions, suggesting that the synergistic effects and interactions among different disconnection core structures merit further investigation.
%\clearpage
%\bibliographystyle{unsrt}
\bibliography{bib/biblio}
\bibliographystyle{elsarticle-num}
%
%\bibliographystyle{unsrtnat}

%%%%%%%%%%%%%%%% END OF MAIN TEXT %%%%%%%%%%%%%%%%%%%%%

\newpage

%%%%%%%%%%%%%%%% START OF SUPPLEMENT %%%%%%%%%%%%%%%

% Figures, tables, equations and pages in the supplement are numbered S1, S2 etc.
\renewcommand{\thefigure}{S\arabic{figure}}
\renewcommand{\thetable}{S\arabic{table}}
\renewcommand{\theequation}{S\arabic{equation}}
\renewcommand{\thepage}{S\arabic{page}}
\setcounter{figure}{0}
\setcounter{table}{0}
\setcounter{equation}{0}
\setcounter{section}{0}
\setcounter{page}{1} % not 0 as \newpage already started a supplementary page
% References continue the numbering from the main text

\begin{center}
\section*{Supplementary Materials for\\ Dual migration modes of unfaulted disconnections on curved twin boundaries}

% Author list for the supplement
Hongrui~He$^{\dagger}$,
Hao~Lyu$^{\dagger\ast}$,
Xueting~Si\\
\small$^\ast$Corresponding author. Email: hao.lyu@dlmu.edu.cn\\
\small$^\dagger$These authors contributed equally to this work.
\end{center}

%%%%%%%%%%%%%%%%%%%%%%%%%%%%%%%%%%%%%%%%%%%%
\section*{Supplementary Information}

\section{Supplementary Note 1}

\begin{figure}[hbt!]
\centering
\includegraphics[width=0.7\textwidth]{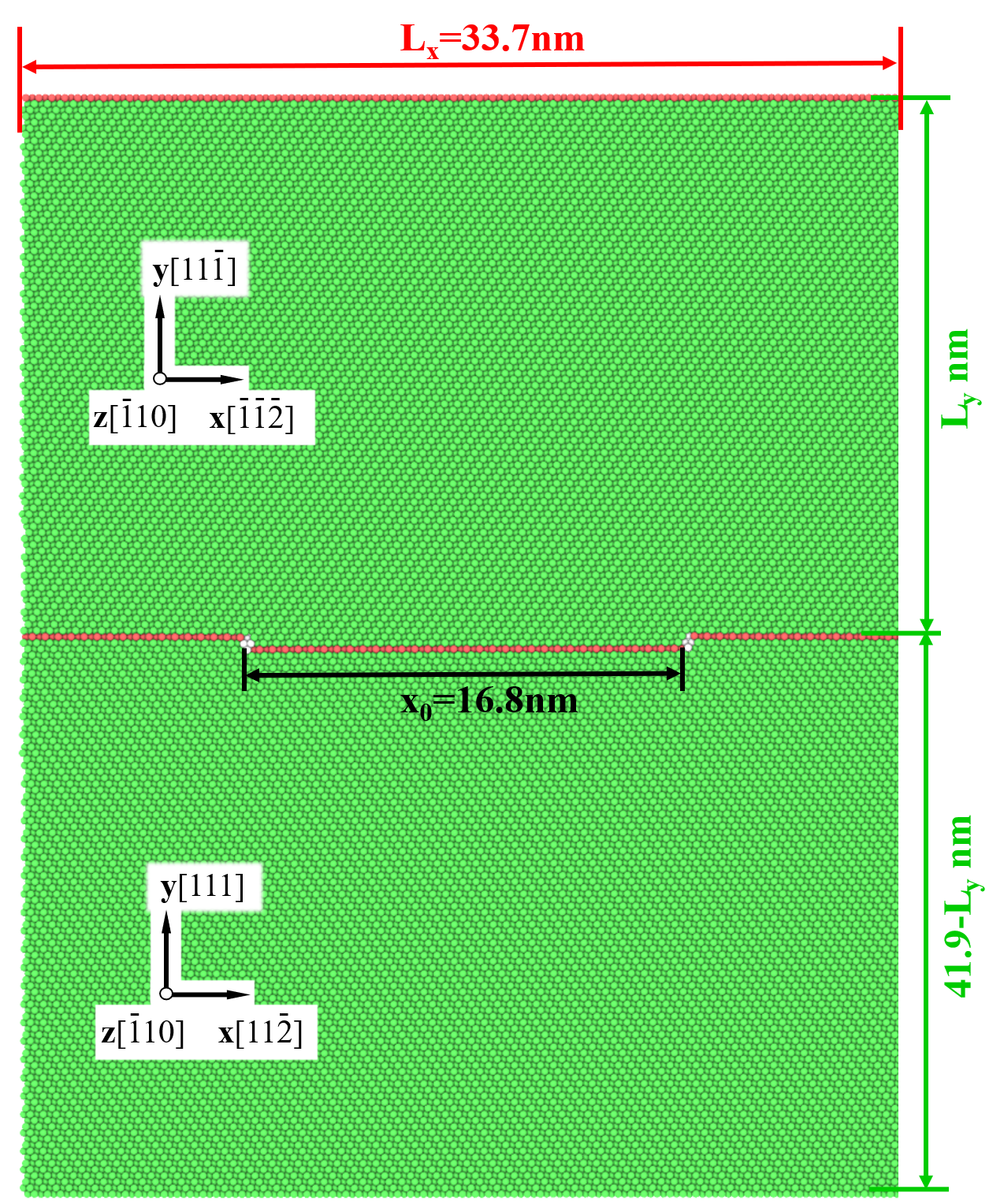}
\caption{Simulation cell for the pristine twin boundary configuration containing a pair of disconnections. Atoms are colored by local crystal structure to distinguish the twin boundary regions from the bulk grains}
\label{S-CNA}
\end{figure}

Supplementary Fig.~\ref{S-CNA} shows the entire simulation cell 
for the pristine twin-boundary configuration with a dimension of 
$\sim 33.7\,\mathrm{nm} \times 42.14\,\mathrm{nm} \times 5.7\,\mathrm{nm}$ 
and contains approximately 489,600 atoms.

%%%%%%%%%%%%%%%%%%%%%%%%%%%%%%%%%%%%%%%%%%%%

%%%%%%%%%%%%%%%%%%%%%%%%%%%%%%%%%%%%%%%%%%%%
The curved interface function is:
\begin{equation}
y=
\begin{cases}
y_{\max}, & x < 8.42 - b \ \text{or}\ x > 25.27 + b \\
y_{\min}, & 8.42 + b < x < 25.27 - b \\
y_{\max} - k(x - 8.42 + b), & 8.42 - b < x < 8.42 + b \\
y_{\max} + k(x - 25.27 - b), & 25.27 - b < x < 25.27 + b
\end{cases}
\end{equation}
\[
y_{\max}=21.07+n d_{111}, \quad 
y_{\min}=y_{\max}-md_{111}, \quad 
k=\sqrt{2},\sqrt{2}/2,\sqrt{5} \quad 
b=\frac{(y_{\max}-y_{\min})}{2k}
\]

where  
$n=-28,0,26$,  
$m=2,3$,  

\begin{table}[htbp]
\centering
\caption{Summary of parameter sets $(n, m, k)$ used for constructing GB configurations.}
\label{tab:GB_parameters}
\begin{tabular}{ccccc}
\toprule
$n$ & $m$ & $k$ & $N_{\mathrm{at}}$ & Corresponding GB type \\
\midrule
$-28$ & $2$ & $\sqrt{2}$ & 489,600     & UFD1 down \\
$26$ & $2$ & $\sqrt{2}$ & 489,600  & UFD1 up\\
$0$   & $2$ & $\sqrt{2}$ & 489,600    & UFD1 mid\\
$0$  & $3$ & $\sqrt{2}/2$  & 489,640   & UFD2 \\
$0$  & $3$ & $\sqrt{5}$  & 489,600   & UFD3 \\
\bottomrule
\end{tabular}
\end{table}
Three distinct grain boundary (GB) configurations were constructed using the same curved surface equation with different parameter sets, as described in Table \ref{tab:GB_parameters}. Varying these parameters modifies the initial atomic stacking arrangement at the interface while preserving the overall boundary geometry.

To systematically explore the configurational space of in-plane stacking states, rigid body translations (RBTs) were applied to each configuration. The upper grain was translated relative to the lower grain along the two in-plane directions (x and z), while the boundary-normal direction was held fixed.

The translational space within one periodic unit of the boundary plane was uniformly discretized into a 10 × 10 grid, yielding 100 distinct translated configurations for each initial GB structure. Following each translation, atoms near the interface with overlapping positions were removed using a cutoff distance defined as a fixed fraction of the nearest-neighbor spacing of the reference lattice.

All configurations were then fully relaxed via conjugate gradient energy minimization under identical convergence criteria, with periodic boundary conditions prescribed in all three directions. The configuration selected for subsequent analysis was not chosen on the basis of minimum grain boundary energy alone. Instead, the optimal configuration was identified according to the following criteria: (i) no stacking fault emission occurs at the interface; (ii) exactly one pair of disconnections is present at the boundary; (iii) the discretized atomic structure faithfully reproduces the prescribed curved grain boundary geometry; and (iv) stable core structure that can be characterized at elevated temperature.
\pagebreak
\section{Supplementary Note 2}
\begin{figure}[hbt!]
\centering
\includegraphics[width=0.8\textwidth]{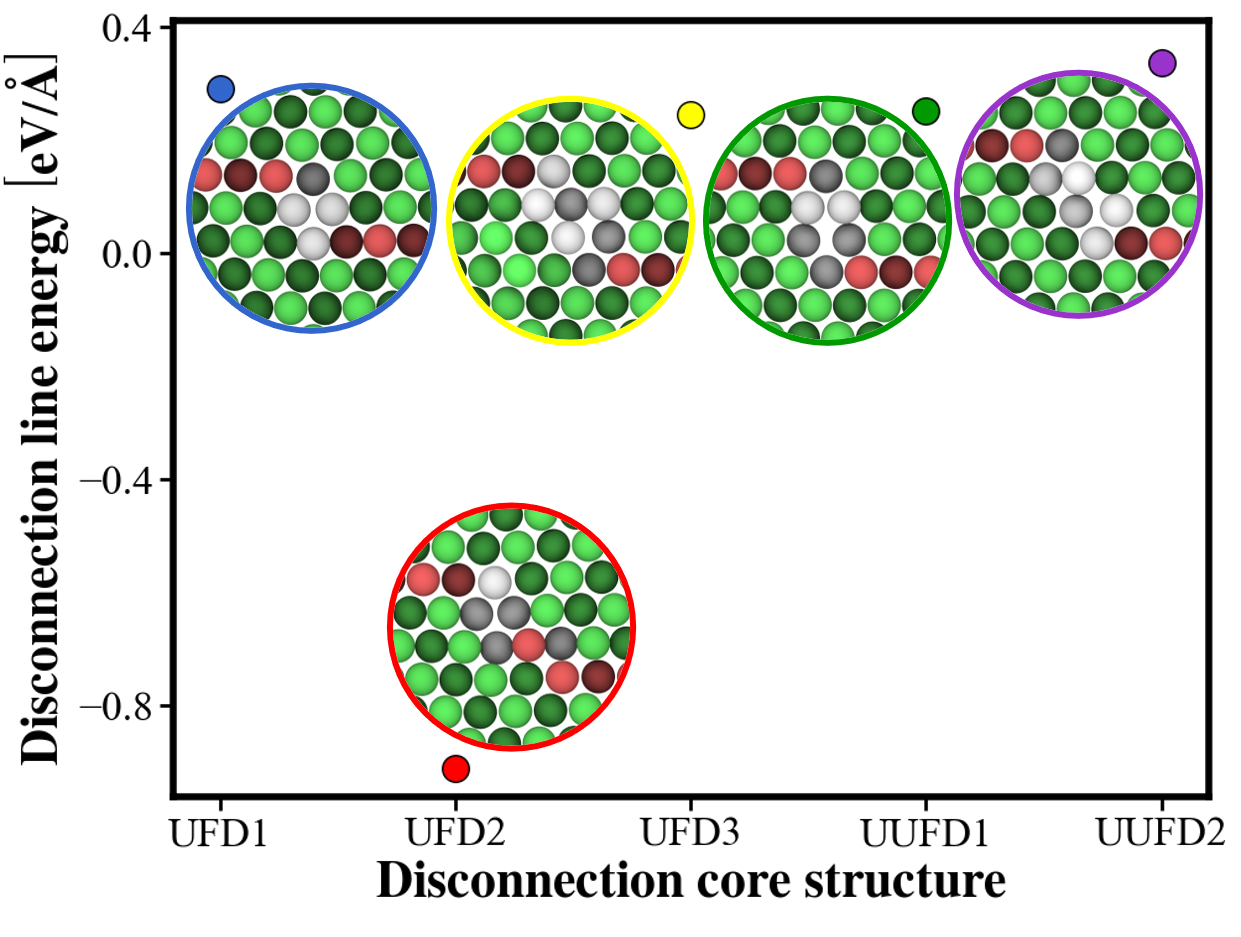}
\caption{Comparison of line energies for different disconnection core structures.}
\label{S-core}
\end{figure}

%%%%%%%%%%%%%%%%%%%%%%%%%%%%%%%%%%%%%%%%%%%%
Supplementary Fig.~\ref{S-core} shows all disconnection core structures identified in this work, each exhibiting a symmetric configuration on both sides of the boundary. Among these, the UUFD1 and UUFD2 structures are metastable and spontaneously transform into the UFD3 configuration upon heating.

The disconnection line energy is calculated as:
\begin{equation}
\Gamma = \frac{E_{\text{tot}} - E_{\text{bulk}} - N_{at} \left( \sqrt{3}a^2/4 \right)\gamma}{L_z}
\end{equation}
where $E_{\text{tot}}$ is the total energy of the system, $E_{\text{bulk}}$ is the total energy of the perfect crystal, $\gamma = 74.65~\mathrm{mJ/m^2}$ is the twin boundary energy, $N_{at}$ is the number of HCP atoms, $a = 4.046~\mathrm{\AA}$ is the lattice parameter, and $L_z$ is the length of the disconnection line.

%%%%%%%%%%%%%%%%%%%%%%%%%%%%%%%%%%%%%%%%%%%%
\pagebreak
\section{Supplementary Note 3}
\begin{figure}[hbt!]
\centering
\includegraphics[width=0.55\textwidth]{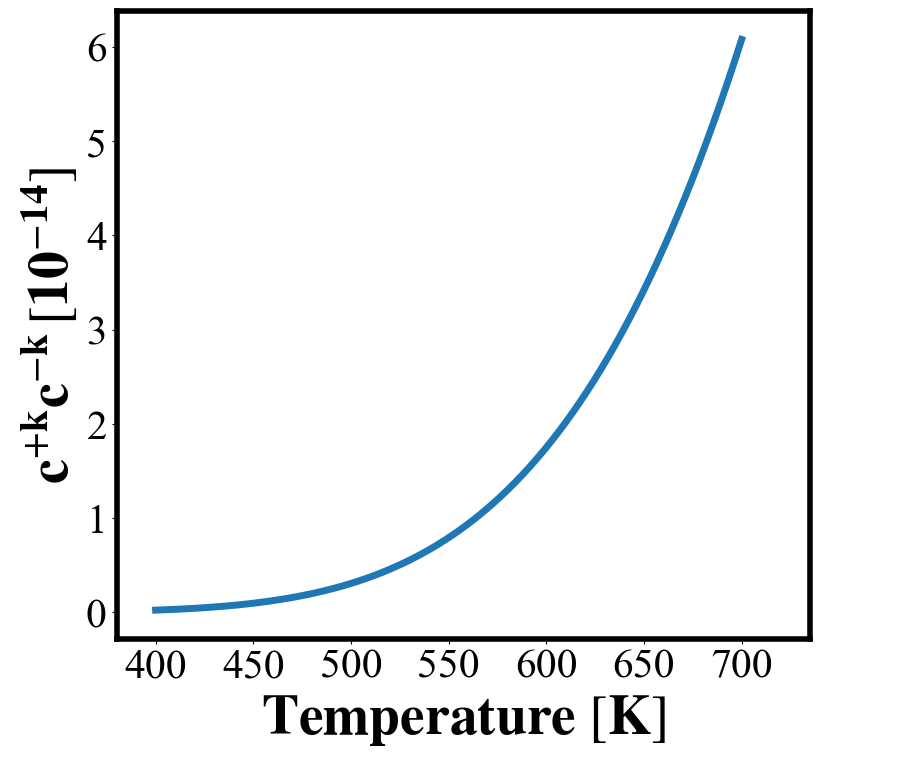}
\caption{Temperature dependence of the activation energies for kink nucleation and kink migration}
\label{S-ck}
\end{figure}

%%%%%%%%%%%%%%%%%%%%%%%%%%%%%%%%%%%%%%%%%%%%
Supplementary Fig.~\ref{S-ck} shows the temperature dependence of the kink-pair nucleation rate,  evaluated using the Arrhenius-type relation: $c_k^+ c_k^- = \frac{1}{a^2}\exp\left(-2F_k/kT\right)$, where the energy barrier $2F_k$ was obtained from NEB calculations, $T$ is the temperature, and $a$ is the interval between adjacent equilibrium kink positions along the dislocation line. The nucleation rates for the kink-pair were then determined based on this thermally activated model.

%%%%%%%%%%%%%%%%%%%%%%%%%%%%%%%%%%%%%%%%%%%%
\pagebreak
\section{Supplementary Note 4}
\begin{figure}[htbp]
\centering
\includegraphics[width=1\textwidth]{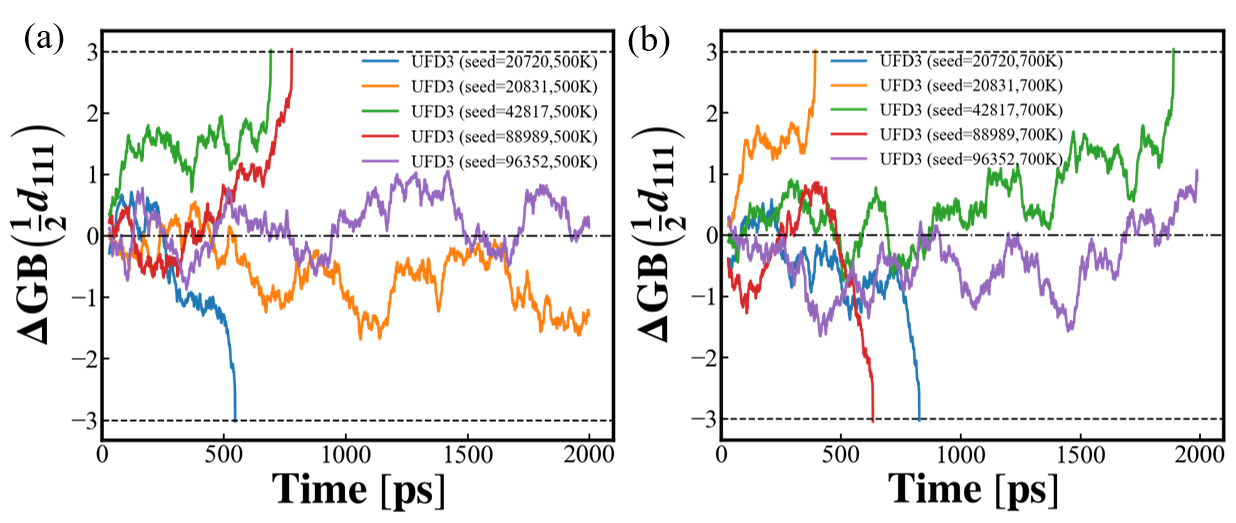}
\caption{Time evolution of grain boundary displacement for the UFD3 configuration at (a) 500K and (b) 700K UFD3-type configurations, obtained from independent simulations with different random velocity seeds. The trajectories reveal stochastic migration behavior across individual runs while consistently showing the tendency of the grain boundary to evolve toward a flattened configuration}
\label{S-UFD3-GB_migration}
\end{figure}

Supplementary Fig.~\ref{S-UFD3-GB_migration} shows time evolution of grain boundary displacement for the UFD3 configuration at (a) 500 K and (b) 700 K obtained from  multiple independent simulations with different random velocity seeds. 

The ensemble of trajectories demonstrates the statistical variability arising from thermal fluctuations; despite stochastic variations among individual runs, the overall migration behavior and trend remain consistent at each temperature.
%%%%%%%%%%%%%%%%%%%%%%%%%%%%%%%%%%%%%%%%%%%%
\pagebreak
\section{Supplementary Note 5}

\begin{figure}[hbt!]%% placement specifier
%% Use \includegraphics command to insert graphic files. Place graphics files in 
    \centering
    %%\hspace*{-1.5cm}
    \includegraphics[width=0.9\textwidth]{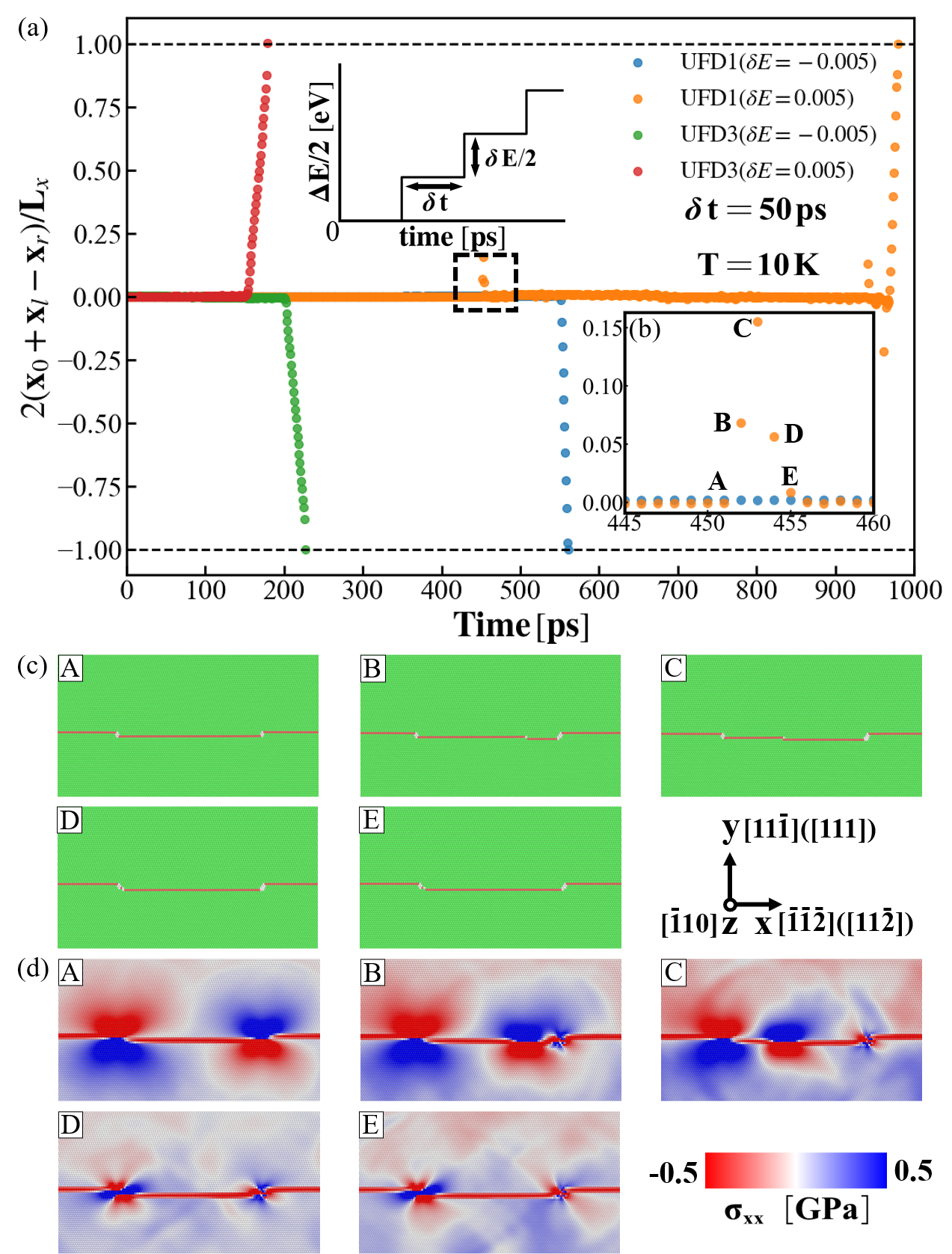}
%% Use \caption command for figure caption and label.
    \caption{(a) Normalized disconnection displacement as a function of time at 10 K for UFD1- and UFD3-type disconnections under positive and negative stepwise driving forces ($\delta$E=$\pm$0.005eV) applied at intervals of $\delta$t=50ps. (b) Magnified view of the boxed region in (a), with labeled snapshots A–E corresponding to successive stages of UFD1 disconnection migration. (c, d) Top-view atomistic configurations and corresponding $\sigma_{xx}$ stress field maps at states A–E, showing the core transformation and stress field evolution of the UFD1-type disconnection under the applied driving force }\label{ECO}
\end{figure}

Supplementary Fig.~\ref{ECO} shows disconnection migration of the UFD1- and UFD3 type disconnection under an efficient calculation of the driving force (ECO) at 10 K.
The figure compares the migration behavior of UFD1- and UFD3-type disconnections under step-wise synthetic driving forces. The results demonstrate that the two disconnection types exhibit distinct critical driving forces required to initiate grain boundary migration. For the UFD1-type disconnection, upward migration is preceded by a core transformation in which a single-layer disconnection nucleates on one side of the boundary and migrates across the interface, where it merges with the pre-existing UFD1-type disconnection to form a UFD2-type disconnection on the opposite side, resulting in asymmetric boundary migration. In contrast, the UFD3-type disconnection exhibits a lower activation threshold and migrates more readily under the applied driving force.
%%%%%%%%%%%%%%%%%%%%%%%%%%%%%%%%%%%%%%%%%%%%
\pagebreak
\section{Supplementary Note 6}

Fig. 7 shows the three configurations (up, mid, and down) were generated using the same surface construction method described in the main text. 
All simulation cell parameters were kept identical across the three configurations, except for the parameter $n$ in the cutting function, where varying $n$ corresponds to discrete changes in the $(111)$ layer spacing.
Specifically, $n = 26$, $0$, and $-28$,$m=2$ were used for the upper, middle, and lower configurations, respectively, yielding $\mathbf{L}_y \approx 24.02$ nm, $20.72$ nm, and $17.90$ nm.
Varying $n$ shifts the vertical position of the twin boundary while preserving the UFD1 step-dislocation pair spacing.
%%%%%%%%%%%%%%%%%%%%%%%%%%%%%%%%%%%%%%%%%%%%%%%%%%%%%%
\pagebreak
\section{Supplementary Note 7}
To ensure the reliability of the calculated minimum energy paths (MEPs), the convergence of the nudged elastic band (NEB) calculations was examined with respect to the number of replicas. As shown in Supplementary Fig.~\ref{S-UFD_N}, NEB calculations performed with N=48, 96, 128 replicas yield nearly identical MEPs profiles for both UFD1 and UFD3 disconnections, corresponding to Fig.~4(a) and Fig.~6(a) in the main text, respectively. The energy barriers and overall path shapes show negligible variation with increasing N, confirming that a discretization of N=48 replicas is sufficient to accurately resolve the reaction pathway. These results confirm that the reported energy barriers and migration pathways are reliable and independent of the replica discretization.

\begin{figure}[htbp]
\centering
\includegraphics[width=1\textwidth]{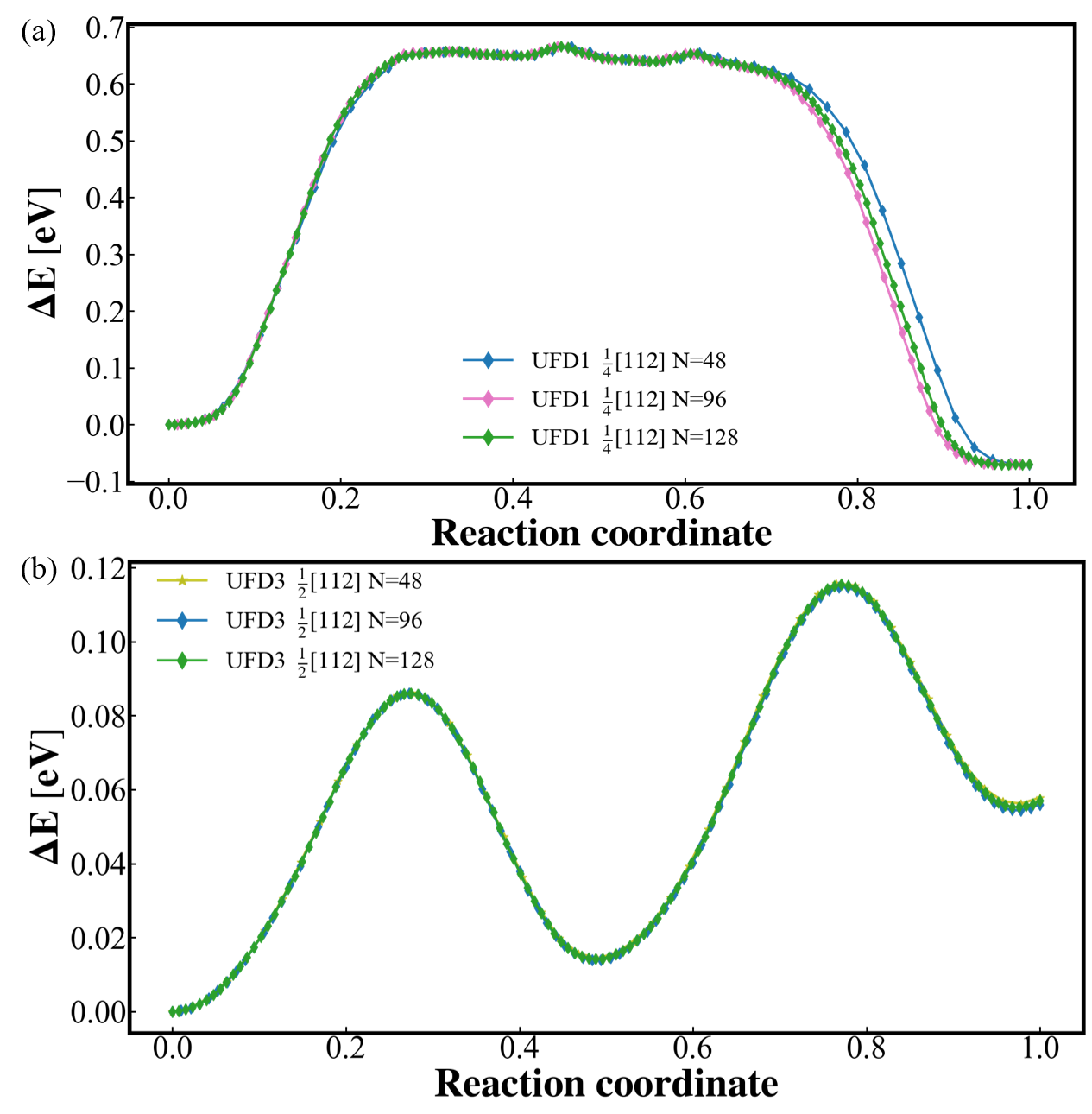}
\caption{Minimum energy paths (MEPs) for the (a) UFD1 and (b) UFD3 disconnections along the $[112]$ directions, respectively, using N=48, 96, and 128 replicas. }
\label{S-UFD_N}
\end{figure}

\begin{figure}[htbp]
\centering
\includegraphics[width=1\textwidth]{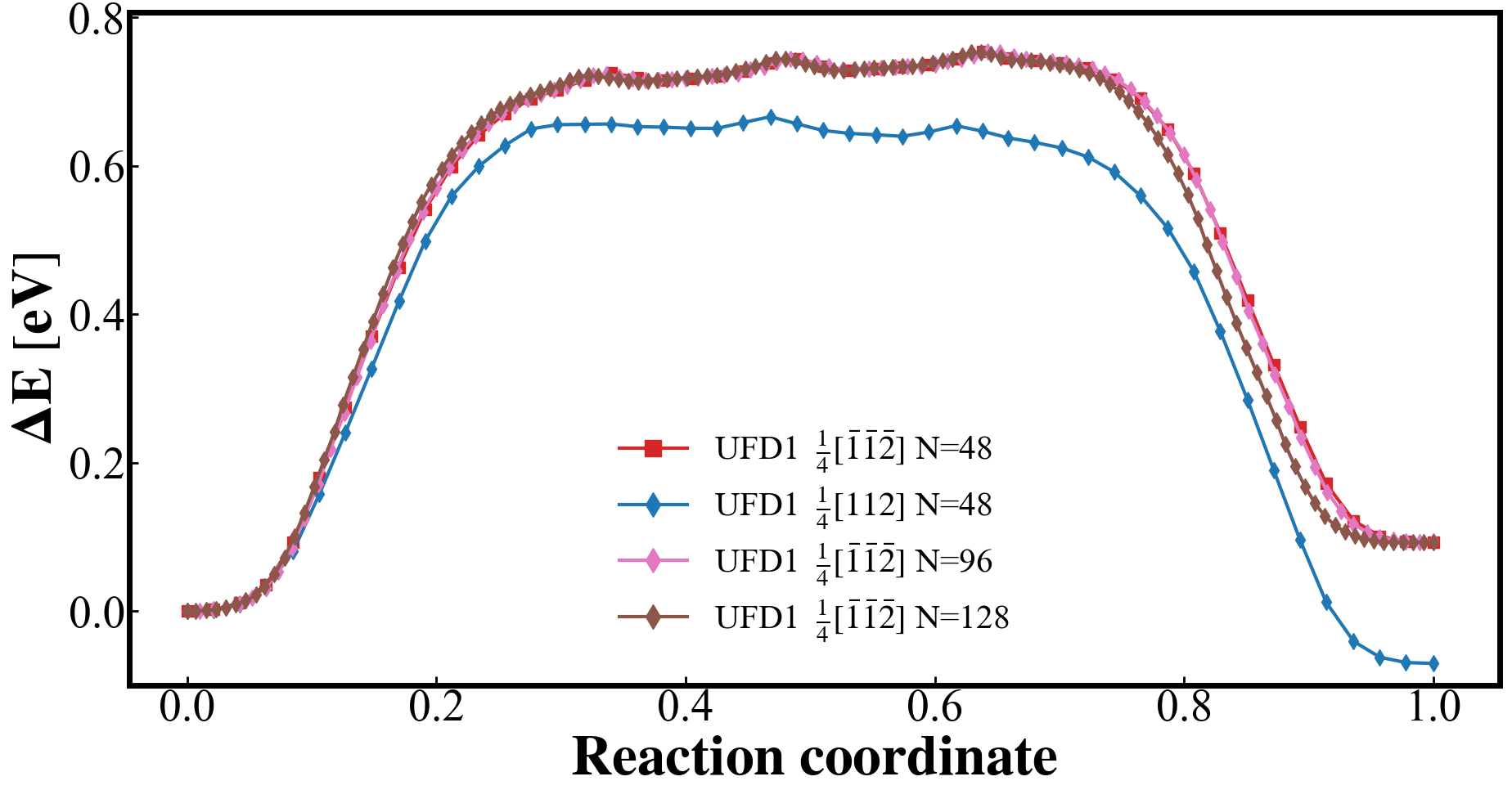}
\caption{NEB energy profiles for the UFD1-type disconnection along the forward 1/4$[112]$ and reverse $1/4[\bar{1}\bar{1}\bar{2}]$ migration directions. The reverse direction is computed with N=48, 96, and 128 replicas, while the forward profile (N=48) is included for reference.}
%\caption{NEB energy profiles for the migration of the UFD1-type disconnection over one lattice translation along the $[112]$ and $[\bar{1}\,\bar{1}\,\bar{2}]$ directions in 600K.}
\label{S-UFD1_neb_inv}
\end{figure}

Supplementary Fig.~\ref{S-UFD1_neb_inv} shows the MEPs for the migration of UFD1- disconnections over one lattice translation along the $1/4[112]$ and $1/4[ \bar{1}\bar{1}\bar{2} ]$ directions, calculated using the NEB method. The energy barriers were determined as the maximum energy along each pathway relative to the initial state, and comparison between the two directions reveals the degree of directional symmetry in the underlying energy landscape.
Replica convergence for the reverse migration of the UFD1-type disconnection was verified using N=48, 96,and 128 replicas.  As shown in Supplementary Fig.~\ref{S-UFD1_neb_inv}, the MEPs obtained with 48 replicas is already well converged, with negligible differences observed relative to higher-replica calculations, confirming that N=48 is sufficient to accurately resolve the energy barrier for the reverse migration pathway.

\begin{figure}[htbp]
\centering
\includegraphics[width=1\textwidth]{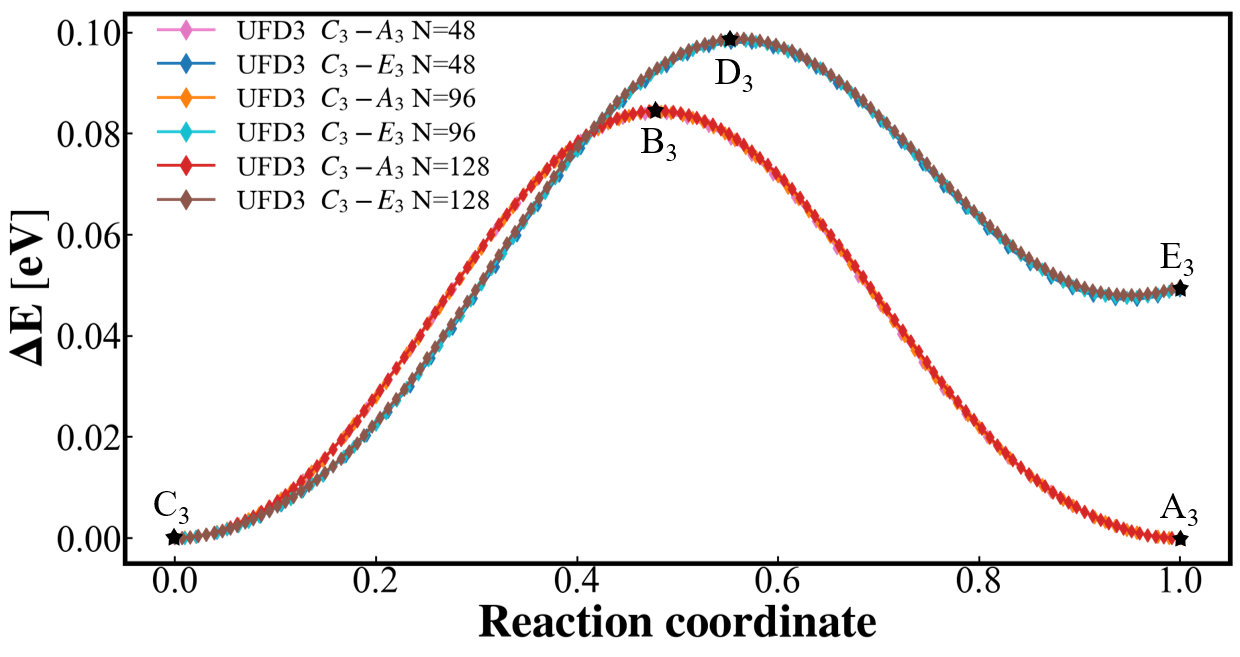}
\caption{
NEB energy profiles for the UFD3-type disconnection for the
C$_3 \rightarrow$A$_3$ and C$_3 \rightarrow$E$_3$ core transformation pathways, calculated using N=48 and 96 replicas}
\label{S-UFD3_neb_inv}
\end{figure}

%%%%%%%%%%%%%%%%%%%%%%%%%%%%%%%%%%%%%%%%%%%%
Supplementary Fig.~\ref{S-UFD3_neb_inv} shows the MEPs for the UFD3 configuration in which the metastable core structure C$_3$ transforms into the stable core structures  A$_3$ or E$_3$ in response to a stable core structure in UFD3 configuration, corresponding to grain boundary migration displacements of 1/4$[112]$ and $1/4[\bar{1}\bar{1}\bar{2}]$, respectively.

\begin{figure}[h!]%% placement specifier
%% Use \includegraphics command to insert graphic files. Place graphics files in 
    \centering
    %%\hspace*{-1.5cm}
    \includegraphics[width=1\textwidth]{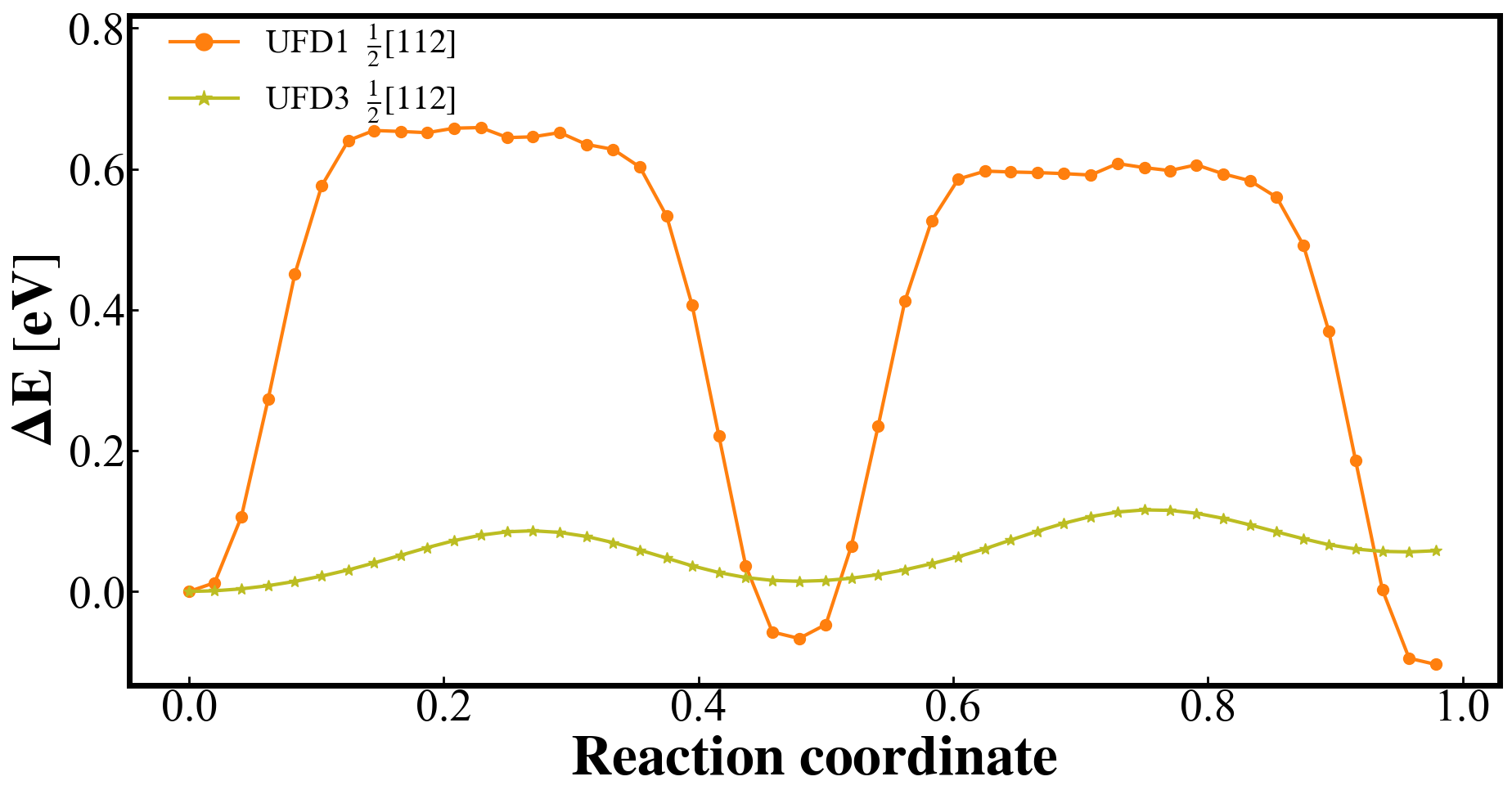}
%% Use \caption command for figure caption and label.
    \caption{NEB energy profiles for the migration of the UFD1- and UFD3-type disconnections over two lattice translations along the 1/2$[112]$ direction in 600K.}\label{S-neb}
\end{figure}

Supplementary Fig.~\ref{S-neb} shows the minimum energy paths calculated using the nudged elastic band (NEB) method for the migration of UFD1-type and UFD3-type steps over a distance of two lattice units along the [112] direction.

\end{document}